\documentclass[review]{elsarticle}

\usepackage{lineno}
\usepackage{graphicx,color}
\usepackage{subcaption}
\usepackage{multirow}
\usepackage{tabularx, tabu, ulem}
\usepackage{booktabs} 
\graphicspath{{Figures/}} 

\usepackage{xcolor}
\usepackage{tcolorbox}
\usepackage{comment}
\usepackage{amsmath}

\modulolinenumbers[5]

\newcommand{\black}[1]{\textcolor{black}{#1}}
\newcommand{\red}[1]{{\textcolor{red}{#1}}}

\makeatletter
\def\ps@pprintTitle{%
 \let\@oddhead\@empty
 \let\@evenhead\@empty
 \def\@oddfoot{}%
 \let\@evenfoot\@oddfoot}
\makeatother
\begin{document}

\begin{frontmatter}

\title{A Survey on Physical Unclonable Function (PUF)-based Security Solutions for Internet of Things}


\author[firstaddress]{Alireza Shamsoshoara\corref{mycorrespondingauthor}}
\cortext[mycorrespondingauthor]{Corresponding author}
\ead{Alireza\_Shamsoshoara@nau.edu}

\author[firstaddress]{Ashwija Korenda}
\author[firstaddress]{Fatemeh Afghah}

\author[secondaddress]{Sherali Zeadally}

\address[firstaddress]{School of Informatics, Computing, and Cyber Systems, Northern Arizona University, Flagstaff, Arizona}
\address[secondaddress]{College of Communication and Information, University of Kentucky, Lexington, Kentucky}

\begin{abstract}
The vast areas of applications for IoTs in future smart cities, smart transportation systems, and so on represent a thriving surface for several security attacks with economic, environmental and societal impacts. This survey paper presents a review of the security challenges of emerging IoT networks and discusses some of the attacks and their countermeasures based on different domains in IoT networks. Most conventional solutions for IoT networks are adopted from communication networks while noting the particular characteristics of IoT networks such as the nodes quantity, heterogeneity, and the limited resources of the nodes, these conventional security methods are not adequate.
One challenge toward utilizing common secret key-based cryptographic methods in large-scale IoTs is the problem of secret key generation, distribution, and storage and protecting these secret keys from physical attacks. Physically unclonable functions (PUFs) can be utilized as a possible hardware remedy for identification and authentication in IoTs. Since PUFs extract the unique hardware characteristics, they potentially offer an affordable and practical solution for secret key generation. However, several barriers limit the PUFs' applications for key generation purposes. \black{We discuss the advantages of PUF-based key generation methods, and we present a survey of state-of-the-art techniques in this domain. We also present a proof-of-concept PUF-based solution for secret key generation using resistive random-access memories (ReRAM) embedded in IoTs.}
\end{abstract}


\end{frontmatter}


\section{Introduction}
\label{sec:Introduction}
\begin{figure}[hbt]
	\centering
	\includegraphics[width=\linewidth,keepaspectratio]{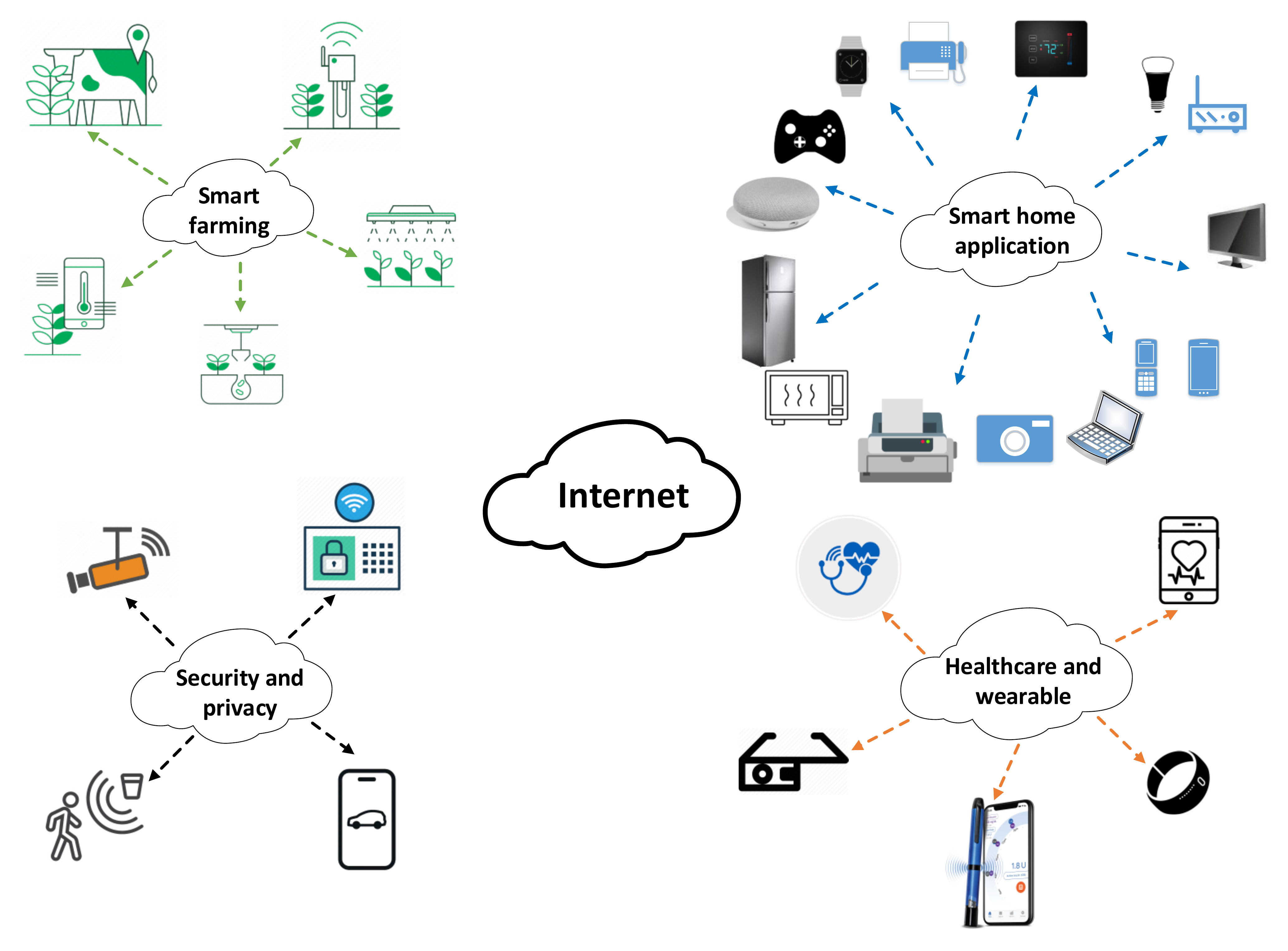}
	\caption{Some examples of IoT applications. }
	\label{fig:Fig1}
   \end{figure}

Wireless communication technologies have managed to imprint \black{themselves} into \black{our daily} lives through \black{the} Internet of Things (IoTs). IoT enables billions of ``things" from tiny sensors to automobiles to interconnect with each other and share their data \black{to create a wide range of value-added services}. IoT systems currently impact different aspects of people\black{'s} daily life. Various kinds of \black{data} including location information, medical information are constantly \black{being collected} by sensors and different electronic and tracking devices \cite{bhayani2016internet}. 
Some examples include using smart watches to set the credit cards with the cell phone's NFC interface to stimulate a transaction protocol and reduce the required time for shopping \cite{galleso2016samsung}; using IoT-based remote health monitoring systems to gather information from patients in a short time and inform the physicians to take timely actions \cite{lu2011application}. Figure \ref{fig:Fig1} demonstrates some applications of IoT systems in our day-to-day life. Fig. \ref{fig:Fig1} categorizes the IoT's applications into different branches such as smart home application, smart farming, security and privacy, and healthcare and wearable devices.
While IoT networks have improved the quality of our life in many levels and opened up a path for several new services, due to the power and computing limitation, high mobility, and the dynamic nature of the network, the security threats and attacks can rapidly propagate throughout the entire network \cite{kloti2013openflow}. One key challenge in IoT networks is the lack of a unified security, identification and authentication standard, while new products and technologies come to market every day without paying enough attention to the potential security threats. Besides the security challenges \cite{zhang2014iot}, there are several other concerns regarding the large scale IoT networks \cite{fuhong2014cooperative} in terms of data fusion \cite{valehi2017graph, valehi2017maximizing} and data management \cite{framling2014universal, shamsoshoara2015enhanced, HanUAVMobility}, complexity \cite{sheng2013survey}, spectrum scarcity \cite{shamsoshoara2019distributed, shamsoshoara2019solution,8406970, afghah2020cooperative, kamhouamodeling, kamhoua2020modeling}, and so on. Moreover, several recent IoT technologies are still based on IPV4 for addressing which compromises the scalability of these networks.


According to Symantec \cite{AsIoTatt94:online}, cyber-criminals progressively target IoT devices since they are developing and expanding rapidly. 
\black{It is estimated that by the end of 2025, 75 billion IoT devices will be connected to the worldwide network \cite{IoTnumb86:online}. In 2019, the attack traffic on IoT network increased by three-fold to 2.9 billion attacks which \black{represents} an increase of 300\% from 2018. \cite{doffman2019}} 

Ransomware is one of the \black{rapidly} growing malwares \black{types}, in which the attacker uses Bitcoin or prepaid credit cards to demand money,
where they do not need to decrypt the public or private keys for stealing cards information \cite{WannaCry56:online, o2012ransomware}. \black{In Ransomware, the attacker prevent the victim from accessing his/her data unless ransom is paid by the victim \cite{brewer2016ransomware}.} For instance, in the earlier months of 2017, \black{many} individuals and businesses around the world  were affected by two huge ransomware attacks, followed by a variant called ``Wanacry" which affected 300,000 computers. 
Petya is another version of the Ransomware attack, which exploited the existence of a third party software to spread and grow itself in the networks \cite{Petyaran66:online,sapienza2017early,yaqoob2017rise}. The prevalence of this malware started from Ukraine, where 12,500 computers were affected. According to Microsoft, this malware has been outspread by using a third-party application \cite{Petyaran66:online}. \black{The Dyn cyber attack used distributed denial of service (DDoS) which to target domain name systems (DNS) which are provided by the Dyn company~\cite{DynAnaly42:online}.} \black{In 2016, } Dyn cyber attack \black{used} DDoS \black{techniques on} several Internet platforms and services in Europe and North America by exploiting \black{security vulnerabilities in} the IoT nodes \cite{dynCyber2016}. 
These attacks have even penetrated into remote health monitoring systems. An article published in the Journal of the American College of Cardiology in February 2018, confirms that cardiac devices can be attacked by hackers leading to severe consequences or even death in some cases \cite{ACC}.


There are several factors contributing to the immense security vulnerability of IoTs including the limited energy available at IoT nodes, their low computational capability, the myriad of available ``things" in a network as well as the heterogeneous nature of the network \cite{samie2016computation}. These characteristics, in particular, the number of connected devices, often result in inefficient performance of conventional security mechanisms. \black{As stated above,} it is anticipated that, about 75 billion IoT devices will be interconnected by the year 2025, which drastically increases the need for advanced security mechanisms for IoT. These ``things" need to transmit the data they collected and intelligently respond and react to the received information. Therefore, it is crucial that information is received from and sent to an authenticated user.

The security challenges in IoTs can be broadly classified into identification, authentication, encryption, confidentially, jamming, cloning, hijacking, and privacy. Encryption has been widely used by several mechanisms in order to send their messages without the risk of being understood by the hackers. \black{Cryptographic methods are a crucial element in securing IoT systems. In theory,} encrypting messages does not allow the hackers to have access to the messages and eliminates the risk of data manipulation. \black{However, encryption alone does not provide or guarantee integrity. For instance, an encrypted message can still be decrypted but the outcome is not fully clear. In addition, encryption by itself cannot avoid malicious third parties from transmitting encrypted packets in the network.} 

\black{Several widely-used encryption} mechanisms includ\black{ing} public Key infrastructure (PKI), 
advanced encryption standard (AES), and elliptic curve cryptography (ECC) rely on \black{secret} keys \cite{dodis2004fuzzy, shamsoshoara2019overview}. In PKI-based systems, there are two sets of keys for each user, private and public keys. The private keys need to be kept secret while the public keys can be known by everyone. In these systems, one of the two keys is used for encrypting  while the other one is used for decrypting. These secret (private) cryptographic keys are expected to not only be reliable, and robust but also perfectly reproducible. Hence, these keys are usually stored in the Non-volatile memory (NVM) of the devices \black{such as ROM and one-time electronic fuse. However,}  due to the 
electrical nature of these memories they are highly susceptible to physical attacks. 
\black{For instance, using a scanning electron microscope (SEM), attackers can implement many invasive threats on these chips. Moreover, using these kinds of memories requires additional fabrication steps during the production of the device. 
Antifuse or electronic fuse is another security technique which is being used for key storage using FinFET transistors \cite{chou201711}. The main benefit of this technology is that the information about the power consumption is not disclosed  during the reading process. Furthermore, this technology enhances the reliability of the read procedure. The  disadvantage is that it fails to remove the key from the device. Once the key is exposed, it cannot be eliminated from the chip.}

The cryptographic keys are sensitive information and therefore, several mechanisms have been developed to protect these keys.
White box cryptography (WBC) is a software based solution to protect these keys and allow secure distribution of valuable information \cite{joye2008white}. WBC requires high processing power and memory and is only applicable to symmetric cryptographic methods; therefore, it will not be a competing candidate for the security of IoT networks. In addition, RSA encryption keys are stored in a specialized chip called trusted platform modules, introduced in \cite{kinney2006trusted}, on an endpoint device (client device) to allow secured encryption.  Physical computing devices called hardware security modules (HSM) were designed to safeguard and manage digital keys. HSM require programming the equipment and interfaces to allow fast data transfer. Key management in \black{an IoT} network is even more difficult due to the increasing number of 
devices. The process of generation, distribution and storage of keys in large-scale IoT networks is \black{still} a major challenge of \black{IoT security}.

 
In general, there are two types of  software-based, and hardware-based mechanisms  to protect IoT devices from various attacks. 
Software-based security mechanisms rely solely on software to protect their messages. They are based on mathematical 
\black{approaches} (e.g., a discrete logarithmic problem) which may not be easily solvable using today's computers but when the existence of quantum computers becomes a reality, they can be solved in a 
\black{shorter time compared to traditional methods} in order to extract the keys \cite{gerjuoy2005shor}. 
Moreover, in software-based security mechanisms, the keys are stored in the NVM of the devices which are prone to attacks. Though software-based security systems were effective all these years, the advancement in hardware and computers may allow the hackers to break them using quantum computers \cite{chen2016report,amy2016estimating,mitchell2019impact,mavroeidis2018impact}.  
As a lot of resources are put towards the creation of quantum computers, their existence will soon be a reality. Therefore, all the existing software security mechanisms are at high risk, which calls for additional security solutions. Hardware based security is one of the possible solution\black{s} to improve the current security mechanisms. 
Hardware-based security uses a dedicated hardware integrated circuit or processor to perform cryptographic functions and store the keys.  They can prevent read-and-write access to data and offer a stronger protection against various attacks. The hardware-based mechanisms such as HSM have been used for crypto processing and strong authentication where it can encrypt, decrypt, store, and manage the digital keys. HSM have been used alongside \black{with} software mechanisms such as PKI, AES to encrypt their messages \cite{barker2011transitions}.

One of the main problems with the hardware-based security solutions is that they are  prone to the Man-in-the-middle attacks. In these attacks, when the hardware security module is stolen, the attackers can clone the device. This can be compared to a simple physical lock and key, where the key is stolen and cloned to mimic the actual key. Physically unclonable functions (PUFs) can provide a  solution to this  mentioned problem. 
Physically unclonable functions were introduced by Gassend et. al. in 2002 \cite{gassend2002silicon} as a security primitive based on hardware. PUF utilizes the intrinsic manufacturing variations in a device to generate a fingerprint of the hardware that offers the valuable advantage of unclonability. This property gives PUFs an edge over other hardware-based security schemes as the hacker cannot clone the intrinsic properties of the device even with physical access to it. Therefore, PUFs are unique to their device and can be used as a security primitive to enable device-based identification, \black{and} authentication. Furthermore, 
PUFs can provide a low cost alternative solution for on-demand generation of  cryptographic keys from the device rather than the conventional methods, where the secret keys are produced and distributed by the server and stored in the IoT device memories {\cite{chatterjee2018building}}. 

The data derived from PUFs is often highly sensitive to environmental changes and the physical conditions where the device is being tested. In other words, the readings from the PUFs are not perfectly reproducible. Therefore, different types of PUFs have been used for the purpose of identification and authentication of devices, where a certain margin of error rate is tolerable. However, even a small amount of variation in the PUF's responses in different conditions can prevent them from being utilized in key generation because the key used for encryption needs to be perfectly reproducible to decrypt the messages. These PUF's responses act as unique fingerprints for the device which are not reproducible.


\subsection{Review of Recent Relevant Survey Papers and the Contributions of this Paper}
\label{subsec:reviewofsurvey}

This survey focuses on the applicability of PUFs-based hardware security for generating keys and authenticating IoT devices. The paper offers a unique and timely survey compared to existing survey papers in the literature by investigating the role of PUFs in IoT security, in particular providing an additional level of security to common key-based cryptographic methods to generate the keys from the devices. 

In 2019, the authors of \cite{mcgrath2019puf} presented a review which discussed several characteristics of PUFs that contribute to their application in authentication followed by a comprehensive classification of different types of PUFs using three different classification systems based on their properties and applications, their parameters and also a chronological classification based on the time these PUF technologies were first introduced. The authors described  relationships between PUF technologies that were not identified previously and investigated other novel forms of PUF which were not exploited. While the mentioned paper offers an inclusive classification of a wide range of different types of PUFs, it does not provide much insight on the challenges of PUFs when it comes to their applications in various security applications. 

In \cite{ehret2019security}, hardware based security techniques in a low-power system-on-chip (SoC) design was surveyed in order to investigate the hardware defenses suitable for it. The authors focus on mitigating the threats faced by the SoC-based embedded and mobile systems as they operate in uncontrolled low power environments.

In \cite{delvaux2015survey}, the authors review the authentication protocols used in 19 different strong PUFs proposed between 2001 and 2014. 
The aforementioned survey reveals different security issues in these protocols and suggests more research is needed on the fundamental physics of a PUF in order to create a truly strong PUF, or else only conventional cryptographic Key generation methods are a promising alternative. The authors also recommend that some of the protocols might leverage the strong PUF to provide side channel attack resistance, but the same physical attack can still be launched on them by adding a machine learning block.

In \cite{babaei2019physical}, the authors state that the compatibility between IoT with limited resources and PUF is its main advantage over other cryptographic solutions proposed for IoT devices. The mentioned paper examines the challenges in utilizing PUF technology in IoT that must be addressed. The paper discusses different threats in using PUF with a focus on the man-in-the-middle and side-channel attacks (invasive, semi-invasive and non-invasive attacks) as well as the defense strategies against these attacks. 
The survey then describes the selection of PUF architectures for IoT based on their robustness to possible attacks, the uniqueness of Challenge Response Pairs (CRPs), and ease of implementation on FPGA. The paper also presents several ways to utilize PUFs to implement cryptographic schemes more efficiently by utilizing them for designing encryption keys, random numbers and electronic signatures.

In 2018, authors of \cite{burg2018wireless} discuss the standards of wireless communication in cyber physical systems (CPS) and IoT and focus on security of these systems. The paper does not study the physical side channel attacks in implementation of security mechanisms for these systems. The paper explains in detail, the various wireless communication standards and protocols. Later, it briefly reviews the security threats in IoT and CPS domains and concludes the paper with recommendations on careful selection of devices from sensors to routers, and auditing the systems using dedicated third party surveillance technologies. The paper does not address the problems in identifying malicious nodes and authenticating known users in the network.  
In \cite{wolf2018safety}, several recent  challenges are identified  as a result of the introduction of IoT and CPS systems.
The authors in \cite{wolf2018safety} focus on different attacks and threats for these systems as well as the challenges related to implementation of CPS and IoT in the wireless network.

In \cite{lin2017survey}, the security  and privacy challenges of IoT in general and possible attacks in different layers of  IoT devices were discussed. In particular, the mentioned paper discusses the security challenges of fog/edge computing-based IoT. In \cite{arshad2018recent}, the use of information-centric networking (ICN) as a possible protocol for addressing IoT devices in terms of in-network caching, content naming schemes, security schemes and mobility handling schemes was introduced. 
The authors of \cite{lin2017survey} stress on the need for ``larger and permanent naming scheme and addressing space for IoT contents and devices"\cite{arshad2018recent}. In \cite{granjal2015security}, the authors focused on securing communications between IoT devices using different protocols and mechanisms and the security weaknesses of IoT at different layers of communication were also discussed.
In \cite{chen2011survey}, the authors discussed the use of programmable hardware such as FPGAs in network infrastructure security. The author highlighted the role of hardware-based mechanisms to address some of the challenges in software-based methods, as well as the potential challenges due to the rising demands of intensive analysis and real time operation for sequential processing. 

\black{In another survey paper \cite{sfyrakis2020survey}, the authors describe how remote attestation schemes can determine the level of integrity of a system and their application in IoT networks, cloud computing infrastructures, and content delivery networks. The authors surveyed hardware-based security devices and cryptographic primitives to achieve security integrity and efficiency and investigated PUFs as a possible solution for remote attestation.}

\black{Another recent survey paper \cite{chowdhury2020physical} studies the application of quantum technology in cryptography and in particular physical security. The authors of the mentioned paper proposed a new type of PUFs called post-quantum PUFs using the fundamental of quantum phenomena.}

The key contribution of this survey compared to the previously published surveys in IoT security is to study the role of memory-based PUFs in authentication and identification of the various IoT devices. More importantly, we discuss the potential advantages and challenges of using PUF-based secret key generation mechanisms to add another level of security to popular key-based cryptographic methods. Such mechanisms, if successful can enhance the security of a huge number of IoT devices against physical attacks.  
The current memory-based PUF technologies do not have the required robustness  to generate fully reproducible responses for low-power IoT devices. This need calls for key generation schemes with error correction mechanisms to generate robust secret keys as required in cryptographic systems as discussed in this paper.

This survey paper is organized as follows.
\black{In Section \ref{sec:Security_Challenges}, we discuss various security challenges in different domains of IoT networks, with a focus on TCP/IP Stack protocol. Moreover, we also briefly describe} different attacks in an IoT network. \black{Next, we} explain the chain of integrated circuit manufacturing and point out hardware attacks based on the vulnerable points. In Section \ref{sec:PUF}, the concept of PUF, their classification along with their application in different security applications are discussed. \black{After explaining different IoT attacks,  Section~\ref{sec:PUF_interaction_security_attack} investigates the role of PUFs in preventing hardware attacks.} 
In Section \ref{sec:FuzzyExtractors}, the role of fuzzy extractors in PUF to generate keys is described. In Section \ref{sec:KeyGenFuzzyExtractors}, we provide a survey on the state-of-the-art key generation mechanisms. \black{Section \ref{sec:FEAttacks} investigates different types of attacks on fuzzy extractors.} In Section \ref{sec:conclusion}, the concluding remarks and future directions of research are discussed.

\subsection{List of acronyms}
\label{subsec:abbreviation}
All of the abbreviations used throughout this paper are summarized in Table~\ref{Tab:Abreviation}. 

\begin{table}
\centering
\caption{Abbreviation table}
\label{Tab:Abreviation}
\resizebox{1.0\linewidth}{!}{
\begin{tabular}{|c|c|c|c|}
\hline
\textbf{\black{Acronyms}} & \textbf{Paraphrase}             & \textbf{\black{Acronyms}} & \textbf{Paraphrase}       
\\ \hline
AES         & Advanced Encryption \black{Standard}       & CPS       & Cyber Physical Systems                
\\ \hline
CRP         & Challenge Response Pair                   & DoS       & Denial of Service                     
\\ \hline
DDoS        & Distributed DoS                           & DES       & Data Encryption System                
\\ \hline
DHT         & Distributed Hash Tag                      & ECC       & Elliptic Curve Cryptography           
\\ \hline
FAR         & False Authentication Rate                 & FE        & Fuzzy Extractor                      
\\ \hline
FPGA        &   Field Programmable Gate Array           & FRR       &  False Rejection Rate                
\\ \hline
HSM         & Hardware Security Modules                 & IC        & Integrated Circuit                    
\\ \hline
ICN         & Information Centric Network               & IoT       & Internet of Things                    
\\ \hline
IP          & Intellectual Property                     & LDPC      & Low Density parity Check              
\\ \hline         
MRAM        & Magnetoresistive  Random Access Memory    & MQTT      & Message Queuing Telemetry Transport   
\\ \hline
NVM         & Non-Volatile Memory                       & OSI       & Open System Interconnection   
\\ \hline
OSN         & Online Social Network                     & PKI       & Public Key Infrastructure     
\\ \hline       
PMKG        & Pattern Matching Key Generators           & PUF       & Physically Unclonable Function
\\ \hline  
Re-RAM      & Resistive Random Access Memory            & RF        & Radio Frequency     
\\ \hline                      
RFID        & Radio Frequency Identification            & \black{RO} 		& \black{Ring Oscillator}   
\\ \hline    
RSA         & Rivest Shamir Aldeman                     & SDA       & Software Defined Network 
\\ \hline        
SEA         & Secure and Efficient Architecture         & SEM       & Scanning Electron Microscope 
\\ \hline   
SMB         & Server Message Block                      & SoC       & System on Chip 
\\ \hline
SOA         & Secure-Oriented Architecture              & SRAM      & Static Random Access Memory       
\\ \hline                                
SS          & Secure Sketch                        		& SSL       & Secure Socket Layer
\\ \hline
TCP         & Transmission Control Protocol             & TSL       & Transport Layer Security               
\\ \hline
ULP         & Ultra Low Power               			& UUID      & Universally Unique IDentifier         
\\ \hline
WBC       & White Box Cryptography						& 			&
\\
\hline
\end{tabular}
}
\end{table}

\section{Security Challenges and Attacks in IoTs}
\label{sec:Security_Challenges}
\black{In this section, we classify the possible attacks on IoT networks from different perspectives and discuss the potential ways the PUF-based security solutions can contribute to mitigating such attacks. Section \ref{sec:Challenges} studies the IoT security challenges in different domains such as data, communication, architecture, and application. Then, in Section \ref{subsec:Attack}, several traditional attacks including denial of service, Sybil are mentioned to familiarize the reader with different attacks and security challenges based on the TCP/IP stack layer. Next, considering the different layers of the IoT's architecture, Section \ref{subsec:ClassIoTAttack} describes a taxonomy of attacks with respect to the IoT structure. Finally, Sections \ref{sec:HardSec} and \ref{subsec:Hardware-Based-Assisted-Security} focus on hardware-based attacks and hardware-based assisted security respectively. The goal of these last two sections is to familiarize the readers with a wider range of hardware-based attacks and possible hardware-based solutions in addition to the PUF-based security solutions which is the main focus of this survey.}

\subsection{Domain Taxonomy to Consider the Security}
\label{sec:Challenges}
In this paper, we address the security issues of IoTs and the potential impact of PUF technology to address some of these challenges.
In IoT networks, the Internet connects  enormous sensors and machines which form a huge network with mobility and heterogeneity characteristics that makes it difficult to protect the network against security attacks. Moreover, the limited energy and computation capability of IoT nodes restrict the utilization of some conventional security mechanisms in these networks \cite{mendez2017internet, alaba2017internet}. One key challenge related to utilizing some traditional security protocols is the heterogeneous nature of IoT networks due to the wide variety of applications of these networks. As a result, different applications require different protection mechanisms. In most scenarios, these applications can be categorized by different characteristics such as user association, openness, and heterogeneity. The heterogeneity of IoTs can degrade the efficiency of cryptographic methods that rely on the key generation and key exchange \cite{zhao2013survey}. Furthermore, noting the characteristics of IoT networks, the security threats such as DoS, routing attack, man-in-the-middle, side channel attack, replay attack, node capture, and mass node authentication are common attacks in these networks.

\begin{figure*}
  \vspace{-5pt}
	\centering
	\includegraphics[width=1.0\textwidth]{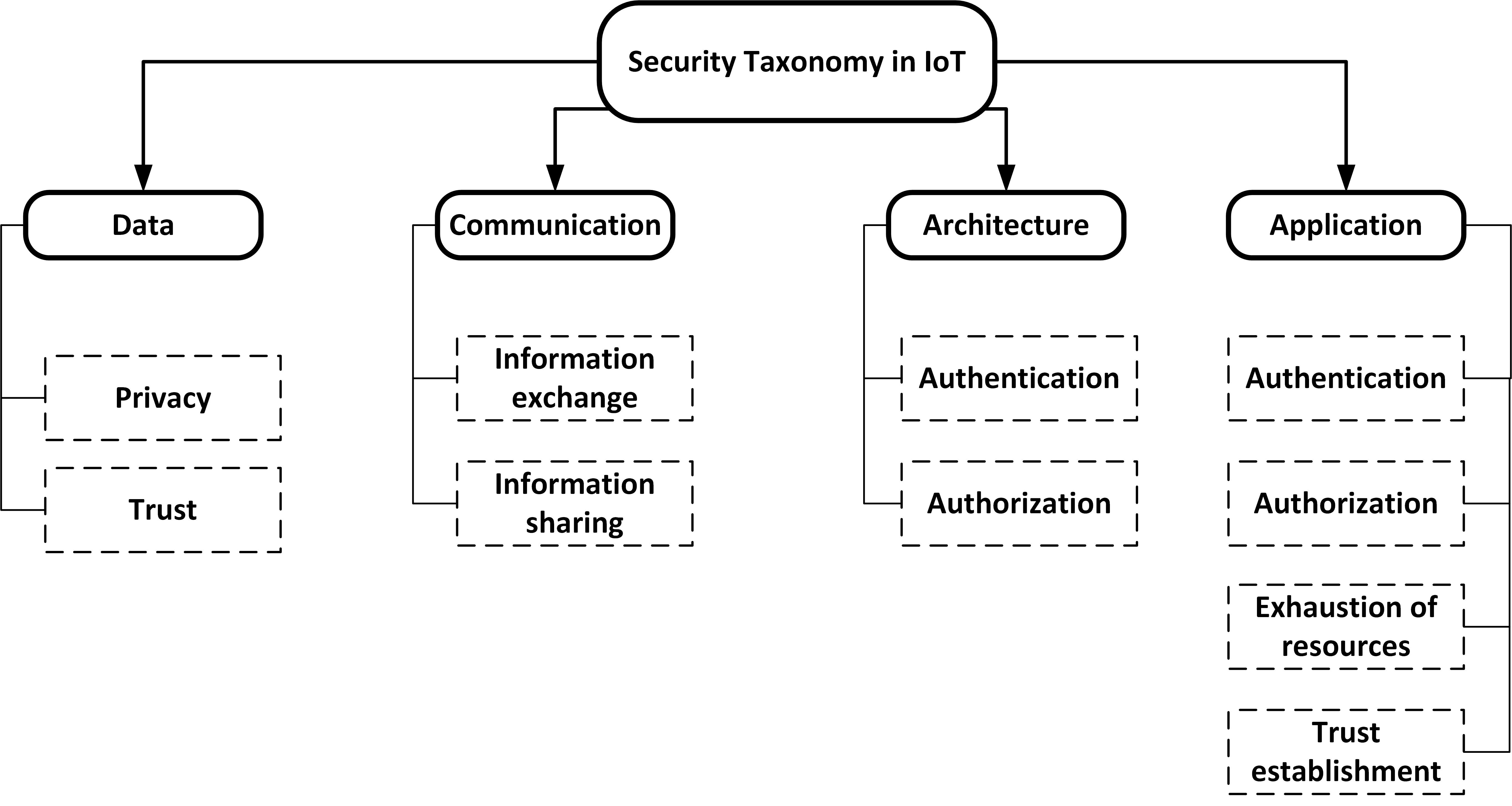}
	\caption{Domains and security concepts in IoT }
	\label{fig:Fig2}
\end{figure*}

Figure \ref{fig:Fig2} shows the various security concepts for IoT based on four domains namely, application, architecture, communication, and data. Based on these stack layers, a security taxonomy can be defined in IoT. \black{We developed Figure \ref{fig:Fig2} based on the discussions in \cite{el2016internet, botta2014integration, hashem2016role} and the literature.}
Next, different concepts of security for each of the application, data, communication, and architecture domains are introduced.

\subsubsection{Data}
Data privacy and confidentiality are important aspects of security \black{in different networks} 
\cite{botta2014integration, keshavarz2018towards, mousavi2019use}. 
In general, \textit{confidentiality} is a security concept which ensures that unauthorized users cannot access the data or 
try to hijack the information. 
\black{Preserving the confidentiality of data is even more challenging in IoTs because of the large number of users and the diversity of network protocols and applications in these networks.} Secure key management is one of the methods which can improve the confidentiality in IoT networks \cite{lin2017survey, capkun2003self}. 
\black{On the other hand, \textit{data privacy} refers to the required regulations related to the collection, storage and sharing of data in such a way to protect the users' personal information (e.g., users' identity) from third parties}.  Most of the time, the main focus of confidentiality is on the encryption of the data; however, privacy defines the level of access 
\black{to} the received data for different users \cite{lin2017survey,zhang2016fakemask}. Finally, \textit{trust} is a concept for the user to accept the security, privacy, and confidentiality in each network. Trust imposes privacy, confidentiality, and security among different layers of IoT, or between different \black{users}, devices and applications \cite{lin2017survey,andrea2015internet,eschenauer2002key, keshavarz2020real}.


\subsubsection{Communication}
Communication in IoT network\black{s} is defined based on exchanging or sharing 
information 
between the users, devices or even exchanging information between different IoT layers. Noting the wide applications of IoT devices in different domains, several communication protocols have been used in IoT networks \black{making} these networks 
vulnerable to various communication attacks \cite{hashem2016role}. As a result, the communication medium is a bottleneck for different attacks such as eavesdropping \cite{pongle2015survey} and Man-in-the-Middle (MitM)\cite{han2015practical}.
\black{Many PUF solutions are available to handle security issues in the communication domain. For instance, in \cite{delavar2017puf}, the authors utilized PUF with an authentication key exchange and a broadcast authentication technique to develop a secure 2-way communication between smart grid meters and the utility infrastructure.} 

\subsubsection{Architecture}
There are no global and specific architectures for IoT networks to validate the security concepts for authorization and authentication. However, various architectures such as software-defined network (SDN) \cite{valdivieso2014sdn}, secure and efficient architecture (SEA) \cite{moosavi2015sea}, smart city \cite{gaur2015smart}, service-oriented architecture (SOA), object security architecture (OSCAR) \cite{vuvcinic2015oscar}, and black SDN \cite{chakrabarty2016secure} are proposed to examine both authentication and authorization.   

\subsubsection{Application}
Scope, scale, 
\black{heterogeneity, accessibility, and repeatability} are \black{among} the application features that can be used to evaluate different security techniques. Trust establishment, exhaustion, authorization, and authentication are considered as different security metrics \cite{gubbi2013internet,rahimi2018security}. Since there is no definite architecture for IoT devices, various techniques have been developed for authentication and authorization in this domain \cite{chen2011novel}.
Noting the wide range of applications of IoT, attacks on these systems can impact several critical domains.

\subsection{Attacks on IoT Devices}
\label{subsec:Attack}
The specific characteristics of IoT devices such as low price, low power, and low computational capability as well as the heterogeneity and large-scale of the network limit the applications of common security mechanisms. Therefore, IoTs are prone to several advanced attacks and security issues \cite{nawir2016internet,li2015internet, trappe2015low} that 
\black{call} for novel security mechanisms in different domains, including identification/authentication, reliability, confidentiality, and non-renunciation. In this section, we review some common attacks on IoTs such as spoofing, altering, replay routing attack, DoS, node capture attack, and Sybil attack.





\subsubsection{Denial of service Attack (DoS)}
In this attack, the attacker attempts to exploit all the reserves and resources in the network which can seriously degrade the network performance. The DoS attack is also called a computational resource attack. These attacks are categorized into two groups: Distributed DoS (DDoS) and individual (single) DoS \cite{alsaadi2015internet,belapurkar2009distributed}. In a single DoS attack, the intruder as a single entity tries to exhaust the resources of the target entity. However, in a DDoS attack, multiple attackers exploit the single entity or a single attacker compromises multiple users to flood the target machine with lots of requests.

\subsubsection{Sybil attack}
Networks with a large number of users are more susceptible to Sybil attacks. In this attack, a single node is identified with different IDs. This means that the unification of entities will be eliminated from the network \cite{sushma2011security}. Based on \cite{zhang2014sybil}, in 2012, 20 million users on Twitter and 76 million users on Facebook were fake. Online social networks (OSNs) such as Facebook, Instagram, or
\black{Twitter} are prone to this kind of attack as they have lots of users. One of the purposes of the Sybil attack is to hijack the information from the OSNs and websites. Since the quantity of IoT sensors and applications is increasing rapidly, they are also vulnerable to Sybil attacks. A Sybil attack can cause users to produce fake and false reports. Users might also receive spam messages from fake profiles and fail to keep their privacy.
Different mechanisms including feature-based mobile Sybil detection, cryptography-based mobile Sybil detection, and friend relationship-based sybil detection (FRSD) are being used to defend IoT networks against Sybil attacks \cite{zhang2014sybil}.


\subsubsection{Spoofed, Alter, or Replay Routing Information}
In these types of attacks, an attacker changes the routing information or tried to manipulate the routing packets by listening to the legitimate transmitter and 
\black{impersonating} the identity of the real transmitter. Then, it sends fake data to the receiver and 
\black{introduces} loops into the network \cite{yilmaz2015survey,sushma2011security}.

\subsubsection{Attacks based on Access-Level}
Based on the level of access to the network, these types of attacks are categorized into two different branches namely, passive and active attacks.

\noindent \textbf{Passive Attacks:}
In most passive attacks, the attacker just eavesdrops the communication between the legitimate transmitter and its receiver to exploit their data. \cite{hossain2015towards,belapurkar2009distributed,alam2014analysis}. 

\noindent \textbf{Active Attacks:}
In active attacks, the intruder attempts to disturb the connection between the legitimate entities, perform impersonation itself, or even disrupt the connection by manipulating the routing information \cite{hossain2015towards,belapurkar2009distributed,mayzaud2016taxonomy,sabeel2013categorized}.

\subsubsection{Attacks in Communication Protocols}
The communication functions of IoT networks are commonly described by the TCP/IP model. Table \ref{table:tab1} presents the taxonomy of attacks that are possible on the TCP/IP protocol stack. 

\begin{table*}[!htbp]
\centering
\caption{Taxonomy of attacks based on different layers of the TCP/IP reference model}
\label{table:tab1}
\resizebox{1.0\linewidth}{!}{
\begin{tabular}{|c|c|c|}
\hline
\rule{0pt}{4ex} \textbf{Layer}
& \textbf{Attacks}     
& \textbf{Attackers' Strategies}                                                
\\ \hline
& {Jamming \cite{mosenia2017comprehensive, li2007optimal, mpitziopoulos2009survey}}                                            

& With radio interference 

\\ \cline{2-3} 
\multirow{-2}{*}{\textbf{Physical}}  & {Tampering  \cite{becher2006tampering, lemke2006embedded}}
& Making fake nodes                                            
\\ \hline
& {Collision \cite{schramm2003new, bogdanov2008multiple, sanadhya2008new}}                                                                                                   & \begin{tabular}[c]{@{}c@{}} Transmit data simultaneously \\  in the same frequency channel\end{tabular}
\\ \cline{2-3} 
& {Exhaustion \cite{tarouco2012internet, hossain2015towards, heer2011security}}             
&                                                    \begin{tabular}[c]{@{}c@{}}                                Multiple collisions and continuous re-transmission \\ until the node runs out of resource     \end{tabular}              
\\ \cline{2-3} 
\multirow{-4}{*}{\textbf{Data Link}} & {Unfairness \cite{varga2017security, burhanuddin2018review}}                     &
\begin{tabular}[c]{@{}c@{}}  Repeatedly ask for the channel to limit others' request
\end{tabular}  
\\ \hline
& {\begin{tabular}[c]{@{}c@{}} Spoofed, or\\  Replayed routing \\ information \cite {liu2012authentication, hossain2015towards, brachmann2012end}\end{tabular}}
& \begin{tabular}[c]{@{}c@{}} Routing loops, changing the source of the\\  route, Repelling network  from selected nodes\end{tabular} \\ \cline{2-3} 
                            
& {Selective forwarding \cite{bysani2011survey, khan2011comprehensive, karlof2003secure}}                                                                                              
& \begin{tabular}[c]{@{}c@{}} Send selected information\\ to the legitimate receiver\end{tabular}                                    \\ \cline{2-3} 
                            
& {SinkHole \cite{krontiris2008launching, choi2009sinkhole, gandhewar2012detection}}                                                                                               
& \begin{tabular}[c]{@{}c@{}} Become the target of all nodes in order to \\ gather all information\end{tabular}                       \\ \cline{2-3} 
                           
& {Sybil \cite{pongle2015survey, borgohain2015survey, zhang2014sybil}}                                                                                                     & 
\begin{tabular}[c]{@{}c@{}}

Create lots of pseudonymous identities \\ to undermine the authorized system
\end{tabular} 
\\ \cline{2-3} 
                            
& 
{Acknowledgement spoofing \cite{senie1998network}}                                                                                                                          
& Spoof the link layer acknowledgement
\\ \cline{2-3} 
& {Hello flood \cite{hamid2006routing, singh2010hello, sharma2010wireless}}                                                                                                                            
& \begin{tabular}[c]{@{}c@{}} Exploit Hello messages to flood the \\ network with these tiny messages\end{tabular}          \\ \cline{2-3} 
\multirow{-11}{*}{\textbf{Network}}   & 
{WormHoles \cite{win2008analysis, jhaveri2010manet}}                                                                                                                     
& Re-transmit data to the IoT nodes

\\ \hline
                            
& {SYN flooding \cite{wood2002denial, yi2005resisting, eddy2007tcp}}                                                                                                                                          
& \begin{tabular}[c]{@{}c@{}} Resend request multiple times to \\ fill the capacity of the transport layer\end{tabular}
\\ \cline{2-3} 
\multirow{-3}{*}{\textbf{Transport}} & {De-synchronization, \cite{joncheray1995simple, roosta2006taxonomy, pathan2006security} }       
& Reinitialize the connection in order to disrupt it
\\ \hline
\textbf{Application}                 
& {\begin{tabular}[c]{@{}c@{}} Reliability attacks:\\  Data aggregation distortion, \\  Selective message \\  Forwarding,\\  Clock skewing \\ \cite{murdoch2006hot, manzo2005time, arackaparambil2010reliability, karlof2003secure, yu2006detecting, roy2006attack, rezvani2015secure} \end{tabular}} & \begin{tabular}[c]{@{}c@{}} Impersonate itself as a reliable node in \\ the IoT network and sends corrupted data\end{tabular}        
\\ \hline
\end{tabular}
}
\end{table*}


\subsubsection{Attacks based on device property}
IoT devices are categorized into two groups: high-end and low-end device classes. According to these types, attacks might have different effects on the devices. They might just result in abnormal behavior or they might stop the devices from working \cite{hossain2015towards}.

\noindent \textbf{High-end device class attacks:}
In this class of attacks, powerful devices such as laptops and computers are used to launch attacks on the IoT network. Most of the time, the Internet protocol is used between the attacker and the IoT network. In these types of attacks, the intruders can use the computing power of CPUs and even GPUs to launch attacks on the IoT network \cite{atamli2014threat,lu2018internet}.

\noindent \textbf{Low-end device class attacks:}
In contrast to the previous class of attacks, in this class, the devices which have low power and energy are engaged in attacks on IoT devices. The attacker uses the radio connection between itself and the IoT device to perform the attack. As an example, smart watches or smart home gadgets are very common in every home. These tiny devices connect to your smart home network which includes TV, refrigerator, cooling system, home security and they can control the configurations of these features. However, these smart home utilities could also be attacked by these little IoT devices \cite{lin2016iot,baccelli2018riot, keshavarz2019automatic}.


\subsubsection{Attacks based on transmitting data}
Sensing and collecting information from the surrounding environment are the main goals for most IoT networks. For this reason, thousands of sensors are being used to gather information. These sensors are also prone to different sorts of attacks which can be used to launch network attacks which can be categorized into six groups such as 1) man-in-the-middle attack, 2)message replay attack, 3) fabrication attack, 4) alteration attack, 5) eavesdropping  attack, and 6) interruption attack.
PUFs can provide lots of solutions for these kind of attacks. For instance, a controlled PUF(C-PUF) can be introduced to handle man-in-the-middle attack. C-PUF is a specific type of PUF that can only be accessed using a specific algorithm which is physically linked to PUF \cite{gassend2002controlled}. The algorithm uses a collision resistant hash function. Thus the only way for the man-in-the-middle attack to be successful is to use the user's program. Therefore, if the user's program has a security leak then the attack can still be successful but not through the PUF itself. Section \ref{sec:PUF_interaction_security_attack} describes the algorithm further. Utilizing this controlling algorithm prevents the man-in-the-middle attack. \cite{gassend2002controlled,tehranipoor2011introduction}.

\subsubsection{Host-Based Attacks}
In this attack, the intruder targets the host resources such as the operating system (OS) or the hardware. The assumption in this attack is that the intruder has managed to access to the host.
Host-based attacks are categorized into three groups: hardware-, software- and user-based attacks. The IoT nodes are usually tiny devices with some applications or software embedded in the OS. The attackers target these three resources of IoT devices and compromise each group with different impact on the overall network \cite{eckmann2002statl}. \black{In Sec. \ref{sec:HardSec}, we focus more on hardware-based attacks and possible hardware-based assisted security solutions}. 




\subsection{Classification of IoT Security Attacks \black{on Different Layers of IoT Networks}}
\label{subsec:ClassIoTAttack}
\begin{figure}[t]
	\centering
	\includegraphics[width=\linewidth,keepaspectratio]{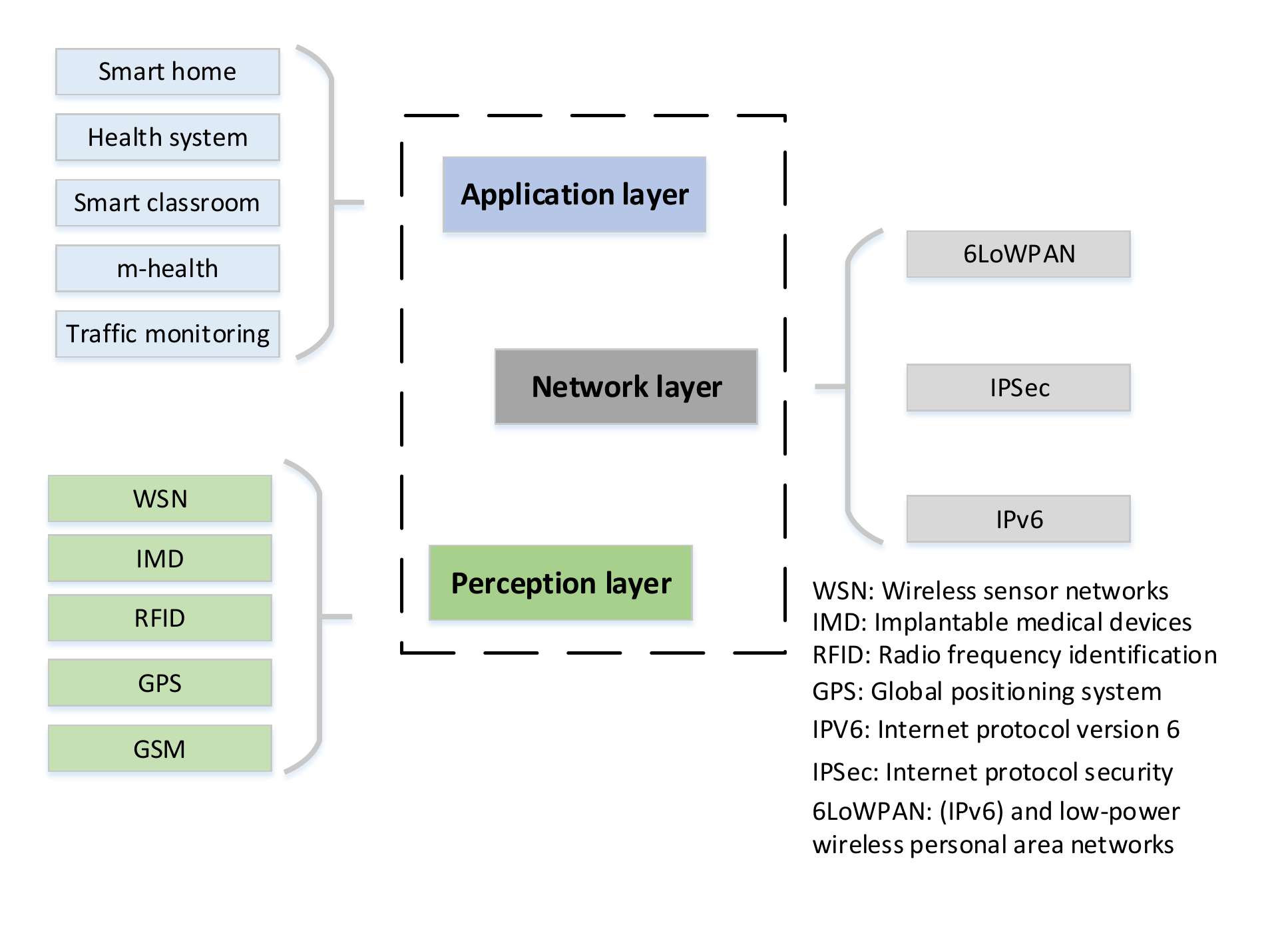}
	\centering\caption{IoT structure}
	\label{fig:IoTStructure}
\end{figure}

This section categorizes common attacks based on the IoT ecosystem. Although there is no well-defined layered model for IoT, Figure \ref{fig:IoTStructure} illustrates a three-layered model for IoT devices \cite{ning2012technology,yang2012multi,song2013security}. These layers include perception, network, and application. First, we describe each layer and then we present a classification of attacks  with respect to the different layers in Table \ref{Tab:TaxAttackIoT}.
\subsubsection{Perception layer}
The perception layer is the lower layer of the IoT networks which handles the interconnection of the nodes in the network. For instance, Arduino boards can use the Ethernet to get access to the Internet, Raspberry Pi can use the Ethernet, WiFi module, or the Bluetooth module to connect to the Internet or other nodes. Each of the communicating devices should have a unique identification called the Universally Unique IDentification (UUID) \cite{leach2005universally}. Most of the time, these IDs are interchangeable. Hence, these UUIDs as System-on-Chip(SoC)s are embedded in the hardware or provided by a secondary chip \cite{song2010semantic}.

\subsubsection{Network layer}
Addressing, network administration, communication channels, and interfaces are the main parts of the network layer. This layer is also responsible for all communications and connectivity for all devices in the network using multiple communication protocols \cite{yang2012multi}. Unlike the Internet, no well-established or standard protocol exists for the network layer in IoT devices. However, Constrained Application Protocol (CoAP) \cite{shelby2014constrained} and Message Queuing Telemetry Transport (MQTT) 3.1 \cite{hunkeler2008mqtt} are two common protocols for the IoT networks. This layer transmits information within the network (other nodes) or outside of the network (e.g., the Internet or a sensor network). Since devices in an IoT network have a limited amount of energy and computation, the role of addressing, forwarding, and routing is pivotal in such networks. 

\subsubsection{Application layer}
This layer makes sure that different entities in the network communicate using the same type of service. This layer is also known as the service-oriented layer \cite{khan2012future} which handles data for different applications based on user requirements and demands. For instance, for applications such as smart transportation, smart home, and eHealth, it can store data into an appropriate database \cite{khan2012future, mousavi2020han}. 


\black{In Table \ref{Tab:TaxAttackIoT}, we provide an overview of some of the main attacks on different layers of the IoT networks. Since the majority of IoT devices have limited on-board power and computation capabilities, existing encryption methods cannot be performed at the device level. } 

Memory-based PUFs have gained high importance in recent years because they are available as embedded memories in every IoT Device as cache or storage, and unlike other PUF technologies, they require minimal or no additional hardware \cite{Koeberl,Holcomb,Sutar,Keller}. Moreover, several memory-based technologies such as memristors  can offer a short process time and a low power supply to  generate the PUF responses which makes them a good security primitive for IoT. The PUFs require a density of 128-256 bits for key generation, which is very small compared to the memory needed in IoT. This will allow us to increase the devices' security as only a small percentage of the entire memory cells will be used for Key generation, therefore identifying those cells will be a challenging task for the hacker. Moreover, by utilizing advanced protocols, which will allow us to change the memory cells we use to extract the PUF response, we can extract many different keys which will increase the security because every time a different key is utilized to authenticate a PUF.
Therefore, memory-based PUFs can offer a unique solution for identification, authentication and even extracting the private cryptographic keys from the embedded memory in these devices without introducing additional fabrication costs to the device.

\begin{table*}[!htbp]
\centering
\caption{Taxonomy of attacks with respect to the different layers in the IoT structure}
\label{Tab:TaxAttackIoT}
\resizebox{1.0\linewidth}{!}{
\begin{tabular}{|c|c|c|c|}
\hline
\textbf{Encryption attack~\cite{andrea2015internet}} & \textbf{Perception attacks~\cite{sonar2014survey}} & \textbf{Network attacks~\cite{kumar2016security}} & \textbf{Application attacks~\cite{nastase2017security}} \\ \hline

\multirow{2}{*}{\begin{tabular}[c]{@{}c@{}}Side channel  attack\end{tabular}} & Node tampering & Sybil attack & Virus and worms \\ \cline{2-4} 
 & RF interference & \begin{tabular}[c]{@{}c@{}}Route information \\ attack\end{tabular} & \begin{tabular}[c]{@{}c@{}}Spyware and \\ adware\end{tabular} \\ \hline
 
\multirow{3}{*}{\begin{tabular}[c]{@{}c@{}}Man-in-the-middle  attack\end{tabular}} & Node jamming & Sinkhole attack & Trojan horse \\ \cline{2-4} 
 & \begin{tabular}[c]{@{}c@{}}Malicious node\\  injection\end{tabular} & RFID spoofing & \multirow{2}{*}{\begin{tabular}[c]{@{}c@{}}Denial of\\ service\end{tabular}} \\ \cline{2-3}
 & Physical damage & RFID cloning &  \\ \hline
 
\multirow{2}{*}{\begin{tabular}[c]{@{}c@{}}Crypto   attacks\end{tabular}} & Social engineering & \begin{tabular}[c]{@{}c@{}}Man in the middle\\ attack\end{tabular} & \multirow{2}{*}{\begin{tabular}[c]{@{}c@{}}Malevolent
\\ \\ script\end{tabular}} \\ \cline{2-3}
 & \begin{tabular}[c]{@{}c@{}}Sleep deprivation \\ attack\end{tabular} & Denial of service &  \\ \hline
 
\end{tabular}
}
\end{table*}

\subsection{Hardware-based Attacks}
\label{sec:HardSec}

\begin{figure}[!htbp]
  \vspace{-5pt}
	\centering
	\includegraphics[width=\linewidth]{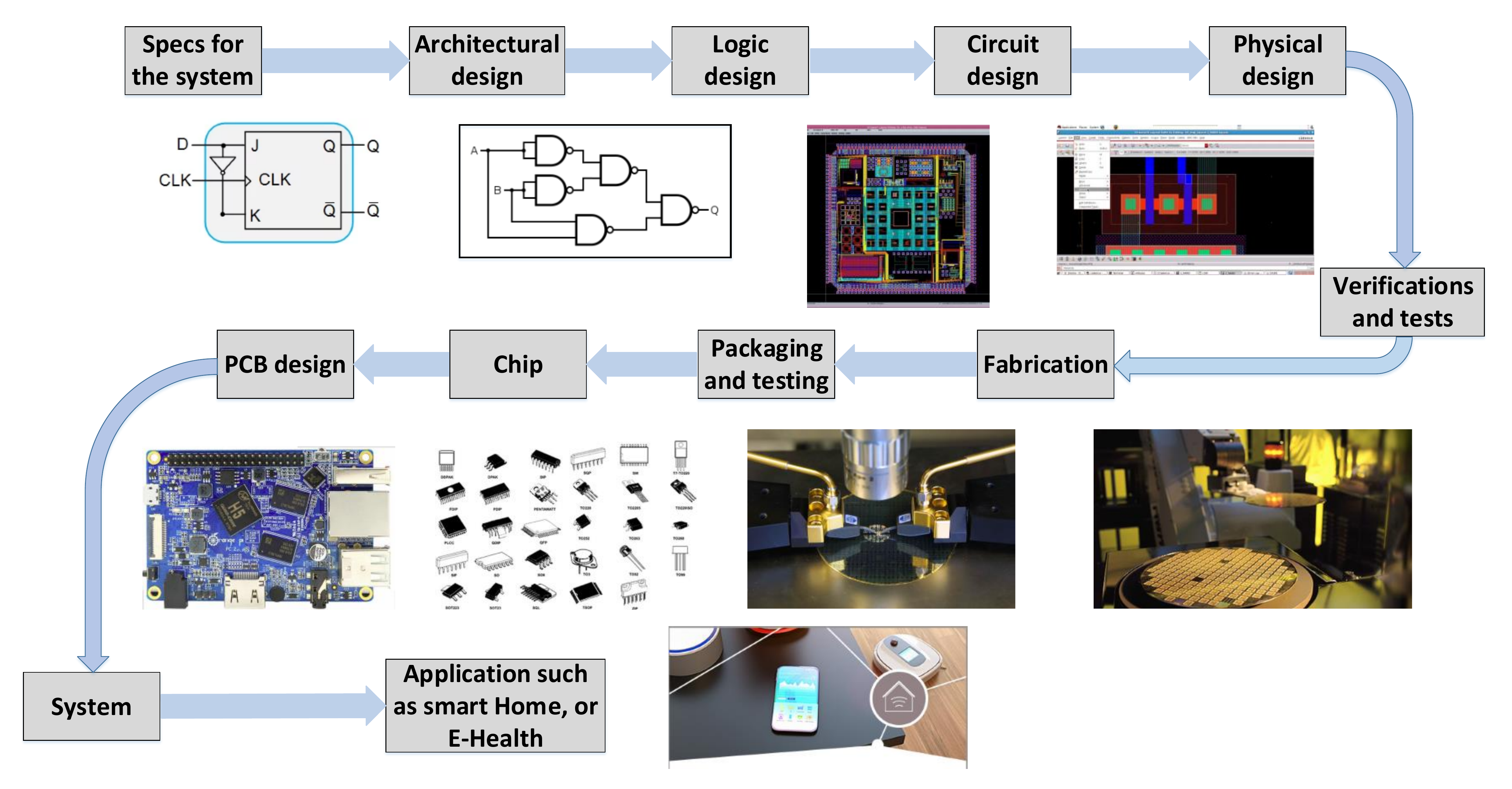}
	\caption{Semiconductor manufacturing process chain for ICs from design home to the application}
	\label{fig:ICFab}
\end{figure}

\black{In this section, we focus on  hardware-based attacks in IoT networks to lay the groundwork to discuss the role of PUFs in securing the IoT devices.}
Based on the production line for integrated circuits (ICs), there are several vulnerabilities based on hardware designs. Figure \ref{fig:ICFab} shows this semiconductor manufacturing process chain which consists of different tasks.  
As a result of the fast and growing tendency in IC industry and production, the global supply line can be targeted and attacked at different vulnerable points. Some common threats faced by the manufacturing process include fake copy, side-channel attack, reverse engineering, intellectual property (IP) hijacking, and hardware Trojans.
Next, we review hardware-based attacks and threats
\black{discussed} in \cite{rostami2014primer,rostami2013hardware}.
\subsubsection{Fake Replica}
In this attack, the intruder counterfeits the original IP illegally. Fake replica and piracy are totally different. Piracy means overbuilding the entire IC. This might happen because the attacker gets access to the design information at different points such as the design or the fabrication. However, a fake replica might happen at different stages such as recycling, packaging, or the new vendor \cite{koushanfar2012can}. Fake replica or counterfeiting can be very harmful to the industry. Since the attacker uses the reputation of the original designer, instead he/she uses expired or old designs to rebuild the ICs or IPs. In most cases, the attacker\black{'s intention is to make profit} 
by selling fake products. However, he/she can also put malicious circuits such as Trojans into those ICs and compromise different critical products and applications such as airplanes, vehicles, drones and UAVs, elevators, and so on.


\subsubsection{Side-Channel Attack}
In some cases, physical states' parameters such as power consumption, timing values, or electromagnetic reflection from hardware can reveal important information to the intruder. In most cases, such information sets can be extracted when the application is being executed where the attacker can perform different tests. Such attacks which involve extracting the behavior of devices \cite{rohatgi2009improved}, are very common in public-crypto systems such as Rivest-Shamir-Adleman (RSA).
RSA uses public and private keys which encrypts and decrypts messages based on modular operations and large exponential values. Two common approaches include calculating the multiplication chain: the first one uses the naive multiplication operation and the second one uses square-and-multiply method \cite{christofpaar2010understanding,mahajan2013study}. In both scenarios, the attacker can use delay analysis to perform a timing side-channel attack. Delay analysis measures the execution time for a number of multiplications which the system uses to calculate the exponential results \cite{RSAcrypt55}. \black{Using the execution time and by exploiting the information regarding the implementation method, the attacker can extract secret information such as the secret private key. For instance, in RSA, the modular exponentiation utilizes the square-and-multiply algorithm to perform multiplication; however, using the statistical analysis and timing analysis with this attack, it is possible to recover the secret key.} In addition to timing side-channel attacks, other attacks such as measuring the photonic emissions, systemic acoustic noise, power consumption, and electromagnetic emissions are common in crypto systems \cite{rohatgi2009electromagnetic,schlosser2013simple,genkin2014rsa}.
\black{One possible solution to mitigate this kind of attack is to use a key-based PUF to extract the key from the device. In \cite{beckmann2009hardware}, the authors proposed a public key exchange method using a PUF which is hard to break by physical and side-channel attacks. Using simulation, they showed that if an attacker uses all available computational resources, then it takes 500 years to break this protocol.}

\subsubsection{Reverse Engineering (RE)}
Reverse engineering is the process wherein the intruder follows a reverse path from the application to the design point for the IC or the IP to reconstruct it, modify it, or implant malicious circuit into it. RE may involve different steps such as i) detecting the technology model which is being used in the design and fabrication  steps \cite{bi2016enhanced}, ii) taking out different parts of design such as gate, logic, circuit, and physical \cite{torrance2011state}, and iii) discovering and observing the functionality of the IP or the IC \cite{rostami2014primer,saeed2017towards}. RE might have different objectives such as hijacking the design, illegally replicating the IC and announcing the technology used in the design. The intruder might use a table of information based on a defined pair of inputs and outputs to evaluate the behavior of the circuit. In this way, the attacker can verify the gate level design from the IP/IC.
\black{The attacker's incentive} might be hijacking at the gate-level, circuit-level, or physical design by performing reverse engineering in order to extract an abstract level of the IP. The attacker can use the abstract level to reproduce the product and to sell it illegally or implant a malicious circuit into the product. 
\black{One solution for mitigating this attack is to use PUF. In \cite{wendt2014hardware}, the authors proposed new approaches using PUFs to obfuscate the hardware. The authors hide the circuit functionality using two methods: i) Hiding the signal path and ii) Replacing a logic using PUFs. They showed that these techniques are resilient to reverse engineering attacks.}

\subsubsection{Intellectual Property (IP) Hijacking}
When the IC is designed, the designers of the IP company or people involved in the fabrication process might hijack the design information without respecting the \black{copyright} terms. Moreover, an attacker at the fabrication stage may reproduce additional chips to sell them on the black market. In these cases, unreliable people can steal the design information and assert a right to possess the proprietary of the IP or the IC \cite{roy2010ending}. \black{One possible solution to protect against the IP hijacking is to use PUFs. For instance, in \cite{anderson2010puf}, the authors used the variation of delays in specific arrays of gates in an FPGA to employ a unique signature for IP protection and anti-hijacking.}

\subsubsection{Trojans in Hardware}
Malicious modifications to an IC can be defined as a hardware Trojan. This Trojan can mislead the communication or cause a failure in control and processing units. In this kind of attack, the intruder can modify and alter the circuit or add a \black{malicious} circuit to it. Since the testing procedures are usually slow and expensive, it is difficult to identify the hardware Trojans after wafer fabrications. Moreover, the technology is merging with Nano- and Pico- meter fabrication design and because of the large space inside of ICs, there are many locations for implanting Trojans. Such locations include different design points such as logic, circuit, and physical and the fabrication process \cite{karri2010trustworthy,tehranipoor2010survey,waksman2011silencing}. \black{In \cite{dupuis2014novel}, the authors proposed novel hardware protection techniques using PUF to prevent the use of hardware Trojan and unauthorized overproduction. The authors minimized the \textit{rare values} in the IC to make it difficult for an attacker to use these values with hardware Trojans.}

\begin{figure}[!htbp]
  \vspace{-5pt}
	\centering
	\includegraphics[width=1\linewidth]{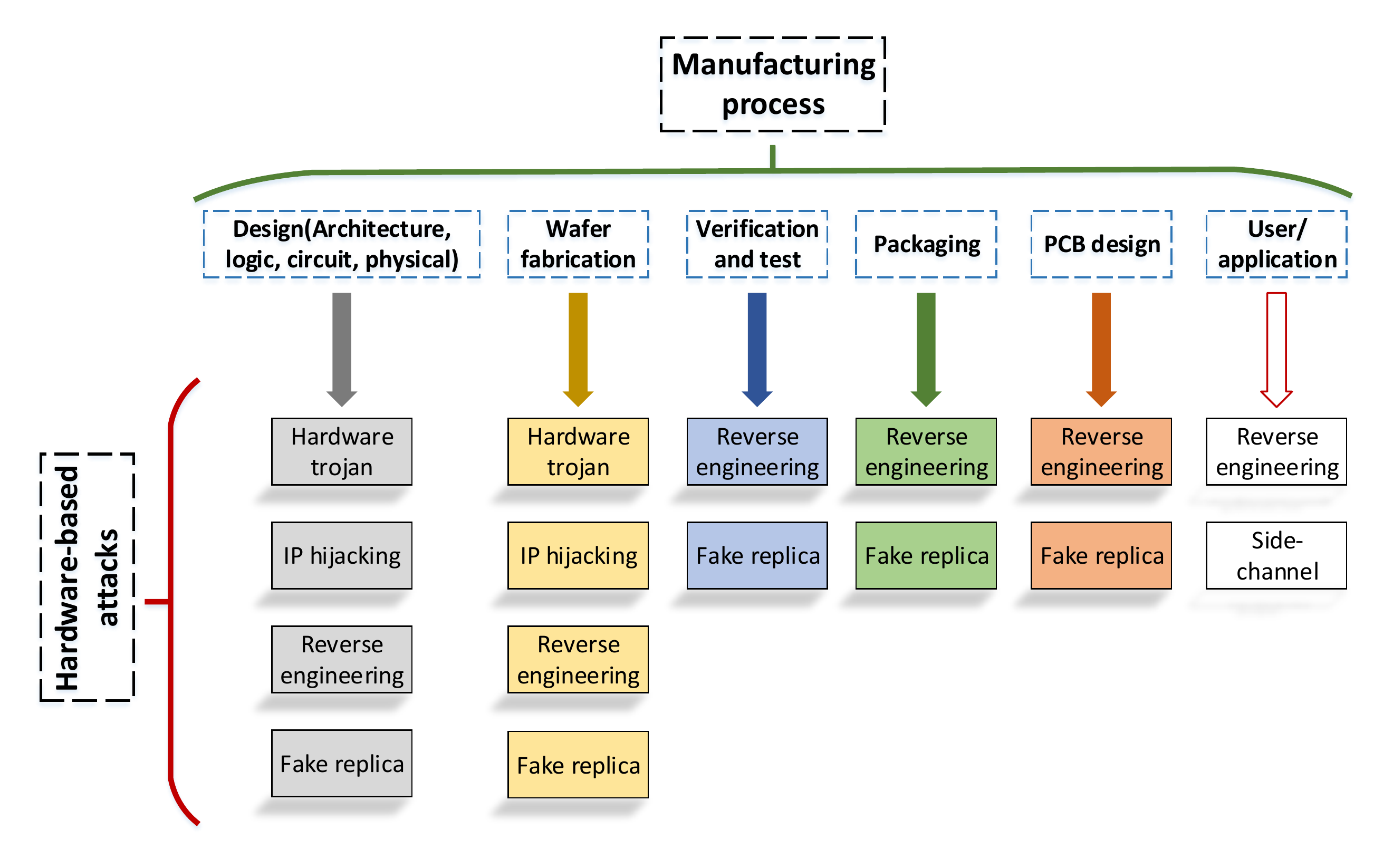}
	\caption{Hardware-based attacks based on different entities in the semiconductor manufacturing process.}
	\label{fig:ICFabattack}
\end{figure}

After introducing hardware-based attacks, Figure \ref{fig:ICFabattack} summarizes these attacks and indicates which entities are vulnerable to specific attacks in the semiconductor manufacturing chain.

\begin{figure}[!htbp]
  \vspace{-5pt}
	\centering
	\includegraphics[width=1\linewidth]{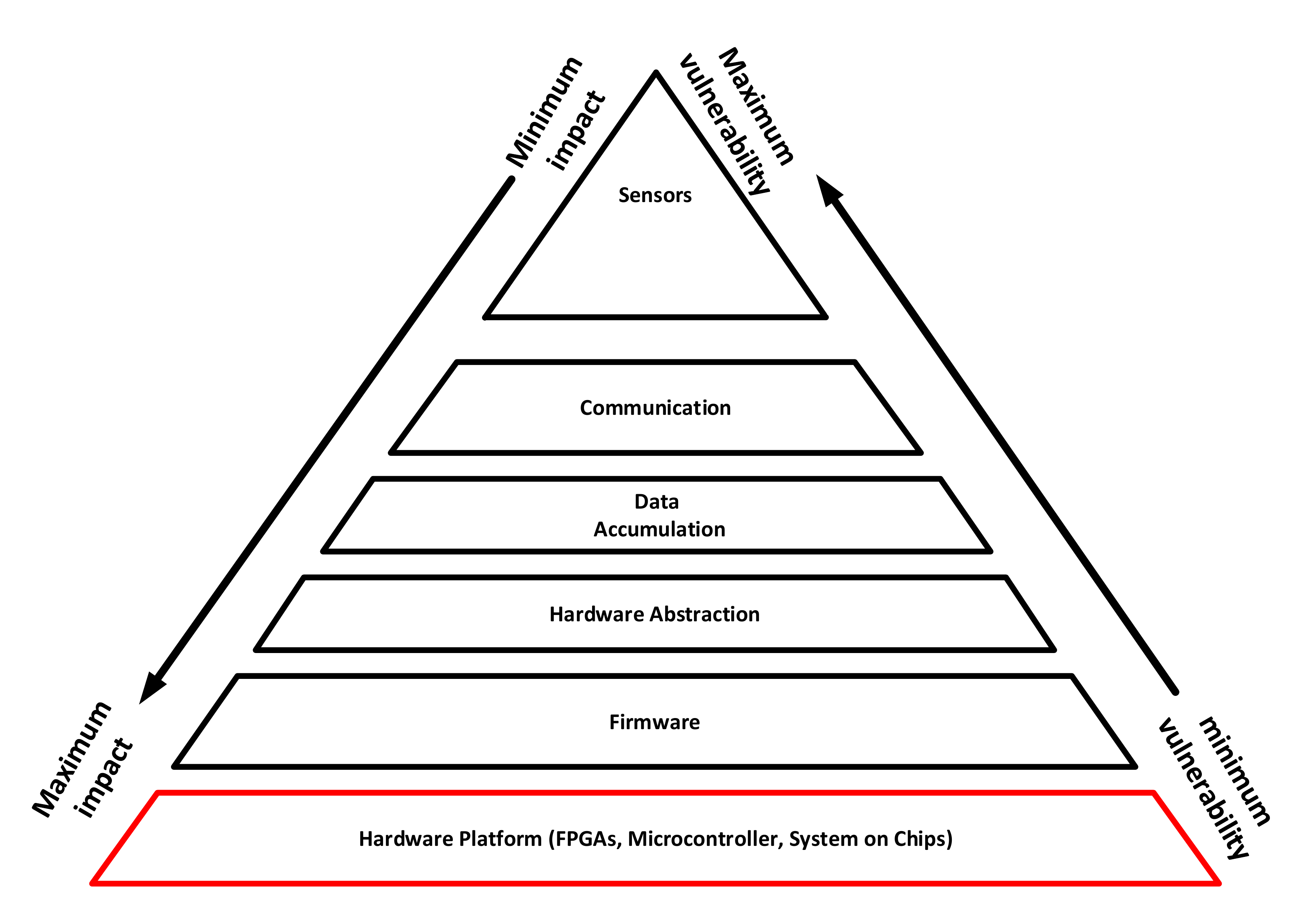}
	\caption{\black{The attack pyramid for IoT devices based on vulnerability and impacts (adopted from \cite{venugopalan2018surveying}).}}
	\label{fig:pyramid}
\end{figure}

\black{These attacks mentioned above are not unique only to IoT devices and they can affect the whole process of IC fabrication. However, considering the  fast growth production rate of IoT devices during recent years, the lack of standards in this domain, and the high demand for low cost devices,  these hardware attacks are more likely to be implemented on IoT devices compared to old-fashioned traditional devices. Another reason is that IoT has the largest application for digital device marketing. Hence compared to other applications, it is more likely to have devastating outcomes on IoT devices because of their mass production. In \cite{venugopalan2018surveying}, the authors present the pyramid of attacks depicted in Fig. \ref{fig:pyramid} which is based on \cite{jimgreen17:online, Microsof90:online}. The peak of the pyramid is the most vulnerable element of the IoT stack with the least impact which are the sensors in IoT networks. The bottom of the pyramid is the hardware platform consisting of the systems on chips, microcontrollers, and Field-programmable gate arrays, which has the most impact on the system in case of any attacks on the IoT systems. In this section, we focus on the manufacturing chain process because it has the most impact in case of attacks.} 

\subsection{Hardware-Based Assisted Security}
\label{subsec:Hardware-Based-Assisted-Security}
In previous sections, an overview of possible attacks on IoT networks was provided. Such attacks can be generally categorized \black{in}to two classes. i) The attacker does not have physical access to the IoT device, hence, the attacker exploits software or network connections to gain access to the IoT device remotely. In this case, the attacker can draw out the cryptographic keys and disturb the authentication mechanism. ii) In the second case, the attacker has physical access to the IoT device or the chip. For instance, the intruder can perform fake replica, reverse engineering or the IP hijacking \cite{lesjak2015hardware}. 
Hence, the existence of an environment is necessary to avoid these kind of adversaries. In the following, two types of hardware-based security methods are proposed. These methods work based on environment splitting which means dividing the the hardware and environment into two sections: i) the secured area, and ii) the unsecured area. In the first approach, which is ``ARM TrustZone", a new state is defined in the processor to bring a meaningful separation. In the latter one, a specific hardware ``Security Controller" such as a microcontroller takes the responsibility to define the reliable environment \cite{trustzone_link}. 

\subsubsection{ARM TrustZone}
\begin{figure}[!htbp]
  \vspace{-5pt}
	\centering
	\includegraphics[width=0.6\linewidth]{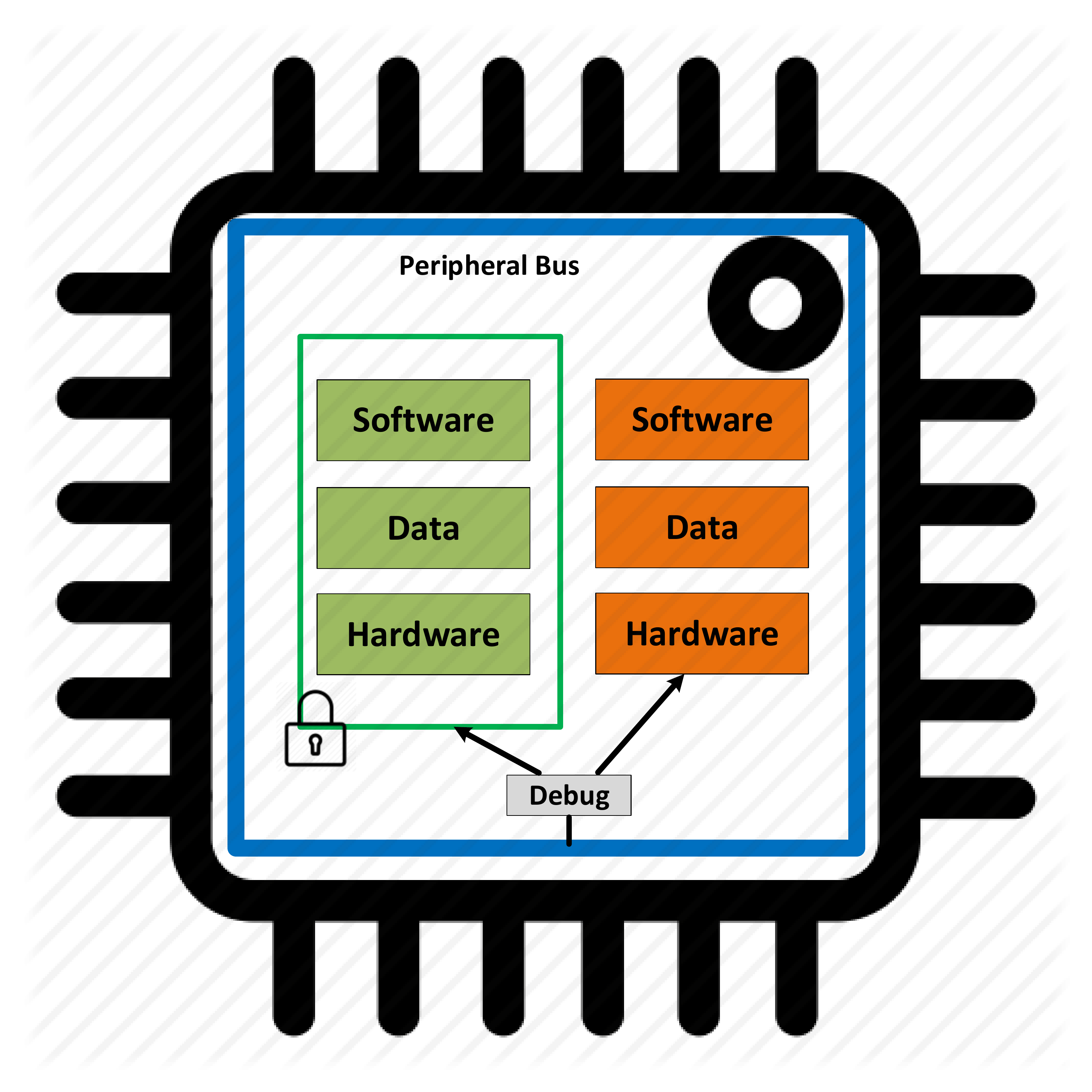}
	\caption{TrustZone concept in ARM based microcontroller. The concept is depicted based on the explanation in \cite{trustzone_link_arm}}
	\label{fig:trustzone}
\end{figure}

This approach is a system-wide method to utilize the security option at the low level for the microcontrollers with the cortex-based cores. \black{Cortex-based cores are a specific family of ARM microcontrollers.} This technology initiates at the hardware level on a single core which divides the processor into two secured and unsecured areas. Since attackers can target the boot up procedure for microcontrollers, this method also secures the boot up process. Core families such as ARM Cortex-A and Cortex-M series support the TrustZone feature \cite{cortex-A_link,cortex-M_link}. The new secure state in the processor splits all partitions in the CPU \cite{trustzone_link}. Using this method, all signals and interrupts of the secured area are isolated from the unsecured one. Figure \ref{fig:trustzone} shows the schematic for this technique. 

\subsubsection{Security Controller}
In the past, several organizations have utilized individual cryptoprocessors which are also known as hardware security modules. Moreover, ATM technologies also utilize those cryptoprocessors as a concept of security in smart cards.
In this approach, the security controller or the secure microcontroller is an individual IC in the IoT device which brings a group of predefined cryptographic tasks. The security controller safeguards the confidentiality and the authenticity of the cryptosystems \cite{lesjak2015hardware}.

PUF is a principle which is being used for authentication and authorization that does not call for any non-volatile memory \cite{gassend2002silicon,ruhrmair2012security}. 
They can be also used for cryptographic key generation, where the digital key is not saved in the device, rather, it is extracted from the physical features of the device. PUFs exploit the random disorder of the physical system in the environment or from the manufacturer. These disorders cannot be re-fabricated again. Hence, they are called intrinsic behavior of the device. Although these features are disadvantageous from the perspective of integrated circuits, they can be advantageous from a security perspective.
In recent years, PUFs face many critical challenges such as reproducibility, wireless transmission of data, and exposure to the predictive models for attacks.

In general, cryptographic methods are currently the most reliable way to secure IoT devices in a vulnerable environment. However, power utilization and key storage are amongst the main concerns when implementing these cryptographic methods in IoT networks.

In the following, Section \ref{sec:PUF} introduces the concept of PUFs and their roles to prevent the aforementioned attacks in this 
section.


\section{Physically Unclonable Functions (PUFs)}
\label{sec:PUF}
\subsection{\black{Introduction to PUFs}}



PUFs use the unique variations introduced in the fabrication of the device, to extract a fingerprint unique to the device. One or more specific parameters of the device such as threshold voltage, critical dimensions etc., are measured when an external stimulus is applied. 
When the devices' parameter is being measured for the first time, the measurement is called an “original response” for a specific input stimulus or a specific address in the memory-called as a “challenge” used to obtain this measurement and they are both stored in the server. When the same parameter is measured again, and the same external stimulus is applied it is called a response.
These challenges and responses form a pair, called the \textit{Challenge Response Pair (CRP)} and are generally compared with each other to validate the identity of the device. The error between the challenge and response of a PUF during the registration and authentication phases is referred to as the \textit{Challenge Response Pair error (CRP error)}.

Subject to the number of possible CRPs a PUF has, they can be broadly classified into: ``strong PUFs'' and ``weak PUFs''. Weak PUFs leverage the manufacturing variability and allow digitization of some ``fingerprint'' of the hardware device. The number of responses in a weak PUF is a function of the number of components in the device used for generation of CRPs \cite{puftutorial}. This fact results in a small number of CRPs with stable responses which are usually robust to environmental conditions. Due to high stability and reproducibility of weak PUF responses, they are generally used for secret key generation. ``Strong PUFs" have a large number of CRPs in a device. Ideally, if the number of unique CRPs is high, even though an attacker gets temporary accesses to the system, he/she will not be able to apply all the responses (brute force attack) and get access to the system. Hence, strong PUFs are generally used for authentication \cite{puftutorial}. However, a large set of PUF responses may offer stronger cryptographic strength as it leads to longer cryptographic keys \cite{maiti2012robust}. Independent CRP refers to the fact that if one CRP is known, one cannot predict the other CRPs in the PUF, hence there is no shared information between two CRPs. \black{PUFs can also be classified based on how their unique-randomness was obtained. If the PUFs had their variation obtained by externally applying additional steps as in the case of coating PUFs they are called \textit{explicit PUFs}. If the randomness was natural through variations in the manufacturing process they are called \textit{implicit PUFs} \cite{mcgrath2019puf}.}

Usability of PUF can be determined by two statistical parameters of \textit{intra-distance} and \textit{inter-distance} which are defined as follows in \cite{bautista2016survey}:
	 \begin{itemize}
	 	\item {\textit{``Intra-distance: the Hamming or the fractional Hamming distance between two different responses to the same PUF challenge"}\cite{bautista2016survey}}
	 	\item {\textit{``Inter-distance: the Hamming or the fractional hamming distance between two responses of two different PUFs to a given challenge."}\cite{bautista2016survey}}
	 \end{itemize}
	 
These measurements indicate the PUFs reproducibility and uniqueness, respectively.

PUF-based security mechanisms depend on the unique CRP's produced from a device\cite{korenda2018secret}. Every PUF device initially needs to be registered with the server in order to use it with any cryptographic method.
During the registration phase, the server uses a stimulus to challenge the client's PUF and as a result a corresponding original response will be produced. This  challenge and response pair is stored in the server's memory. During the authentication process, the server uses the same challenge for the client's PUF to extract the corresponding response. These responses depend on the manufacturing behavior and variations in PUF.
The user is authenticated if the number of bits in error between a CRP at the registration phase and authentication phase is statistically low enough \cite{cambouafghah2015, ruhrmair2010strong}.


Another application of PUFs is to utilize the high randomness introduced during its manufacturing to create a secure key from the device. Such key generation requires ideal PUFs that are robust, tamper evident, and unpredictable. In order to correct the noise in a PUF response and generate cryptographic keys, the concepts of a secure sketch in a fuzzy extractor which are described in sections \ref{sec:FuzzyExtractors} and \ref{sec:KeyGenFuzzyExtractors}, are utilized. Secure sketches use the concept of error correction coding to ensure that we recover the original PUF data from the noisy PUF.

\subsection{Types of PUFs}
Different components can be utilized to extract fingerprints from a device. Initially randomness was physically introduced into a device to extract a fingerprint, whose examples are optical and coating PUFs. An optical PUF uses the physical property of a transparent material, in which the light particles scatter in an uncontrolled manner. When a laser beam falls on it, a unique and random pattern is produced \cite{pappu2002physical}. A coating PUF can be built by filling the space between a network of metal wires on top of an IC with a randomly doped opaque with dielectric particles \cite{puftutorial}. Due to the random placement of doping, each couple of wires will have a random capacitance value. This PUF is generally used on the top layer of the ICs which is generally used to protect the underlying circuits from attackers' inspection. The capacitance between the wires will change when a portion of the coating is removed. These PUFs have been used as RFID tags \cite{tuyls2006rfid}.

Current technologies prefer to utilize the PUFs designed based on intrinsic variations, because they are already embedded in the device. \textit{Silicon PUFs} exploit the intrinsic variations in the IC manufacturing process. \textit{Leaked current-based PUFs} were dependant on the concept that the combination of different intrinsic variations in a circuit will result in a different leakage current. Another example of silicon PUFs are \textit{Delay-based PUFs}, where distinct delays are caused in a circuit due to the manufacturing variations in its components even in an identical layout. The most popular examples of delay-based PUFs are \textit{Arbiter PUFs} and \textit{Ring Oscillator PUFs} \cite{shamsoshoara2019ring}. These PUFs need huge groups of device components  to make them secure. These PUFs tend to take a substantial amount of chip space and are vulnerable to side channel attacks because they give off information due to heat and therefore, they may not be suitable for IoT nodes.
\black{Therefore, PUFs which can be easily deployed, occupy less space, and require less power are required for security purposes. These characteristics can be found in PUFs made from memory devices.}

\subsubsection{Memory based PUF's}
Memory-based PUFs utilize the memory chips readily available in different devices and hence can be easily deployed in any device, to allow PUF based security in a network. PUFs can be made from different types of memory including SRAM, Flash, MRAM, memristor, and ReRAM. 

\black{SRAM cells are made of cross-coupled inverters which are connected by access transistors and because of the intrinsic manufacturing variations, SRAM cells typically settle into a ``0" or ``1" state consistently \cite{vijayakumar2017improving,bohm2011microcontroller}. SRAM PUFs were introduced in \cite{holcomb2007initial} where the initial values of the cells, on powering on the SRAM were used to generate a unique fingerprint. SRAMs tend to emit energy when they switch states which can be detected by checking the wavelength of the laser by using a signal analyzer. When this side channel information is leaked, it can provide enough information to the attacker about the device in order to clone it \cite{ReRAMternaryPUF,helfmeier2014physical}. }

In our experiments, we utilize addressable PUF generator protocol to extract the fingerprint of the Memory device \cite{cambou2019encoding}. In this protocol, a random number and a password which is known to both the client and server is exclusively- or'ed and sent to a hash function. This random number acts as our challenge for the protocol. The message digest obtained from the hash gives us the information in terms of address of the memory cell, from which the response of the PUF is extracted. This protocol can also be further extended to use multiple addresses to extract a key from different places of the device. A new key can also be easily developed by changing the Random number which will give rise to new addresses for fingerprint extraction.

\black{
The manufacturing of a Re-RAM is very similar to memories which use CMOS technology and hence can be easily integrated into the IC. Voltage is used to program Re-RAM and erase cells while current is used to read the resistance values of the cells \cite{pavan1997flash}. Only specific parameters such as low and high state resistance, which are used in making the CRPs in the device can be measured consistently.}

\black{Re-RAM-based PUFs use the value of $V_{\text{set}}$ after programming or the resistance of each cell, to differentiate between ``0" and ``1" states. The flipping probability of cells during response generation is higher when their resistance values are closer to the threshold, when they are subjected to voltage changes, temperature, aging or electromagnetic interference. This could result in 5-20\% CRP matching error rate if the number of cells whose resistance is close threshold is high. 
}

PUFs designed using Re-RAMs are immune to side channel attacks without direct access to the chip as they work at or below noise level. Re-RAM and magnetoresistive random-access memory (MRAM) offer low power options when compared to current Flash technologies because they rely on resistance. Table \ref{opReq} compares the operation requirements for Flash, Re-RAM and MRAM.

 \begin{table*}[t]
\centering
\caption{Operation requirements for current and novel memory technologies\cite{chang2007endurance}\cite{akinaga2010resistive}\cite{tehrani2006status}}
\label{opReq}
\resizebox{1.0\linewidth}{!}{
\begin{tabu} to \textwidth{|c|c|c|c|}
\hline
Operation              & Flash                        & ReRAM           & MRAM            \\ \hline
Program parameter      & NOR Vds = 5V; NAND Vgb = 15V & Vset = +100mV   & Current: 500uA  \\ \hline
Program power required & 1mJ/bit                      & 10pJ/bit        & 100pJ/bit       \\ \hline
Program speed (ns)     & 5000ns/block                 & 2-20ns          & 2-20ns          \\ \hline
Read parameter         & Voltage: 10mV                & Current: 1-20uA & Current: 1-20uA \\ \hline
Read power required    & 10 pJ                        & 1pJ             & 1pJ             \\ \hline
Read speed (ns)        & 50ns                         & 2-20ns          & 2-20ns          \\ \hline

\end{tabu}
}
\end{table*}

\subsection{Comparison and Applications of Different Types of PUFs}
Table \ref{tab:comparePUF2} demonstrates different PUFs based on their types and their specific features. 
Since the goal of PUFs is to enhance the security of entities and nodes in a network, they can utilize the physical and application to enhance the security. We can also develop a cross-layer design which considers PUF along with these two layers. However, there is no need for any additional implementation, it is possible to use the built-in structures and element to enhance the security for the aforementioned layers (physical and application). Based on the security challenges mentioned in sections \ref{subsec:Attack} and \ref{subsec:ClassIoTAttack}, PUFs are introduced to address several security issues of IoT networks. In most scenarios, PUFs are used to achieve authentication and authorization.

\black{In \cite{puf-cprng2020}, PUF was utilized to create chaotic Pseudo random number generators (random number generators which are extremely sensitive to initial seed), and the PUF was utilized to create the initial seed. PUFs were also used in \cite{wang2019puf}, to enhance the strength of the password generated by improving the entropy by using PUF as an entropy pump which showed a 48\% improvement in entropy from human passwords. }

\black{In \cite{cambou2018addressable}, PUFs were part of the PKI architecture proposed which did not require storage of reference patterns and passwords in order to mitigate attacks related to stolen databases. PUFs were also a part of digital signature schemes proposed in \cite{cambou2020secure}. In order to verify the digital signature, these schemes require error correction mechanisms to obtain a stable response from the PUF.  PUFs are also deemed one of the candidates for a true random number generator in \cite{cambou2019data}, where the random number was generated using a plurality of cells from the PUF response. }

\begin{table}[bt]
\caption{Comparison of different types of PUFs.}
\centering{
    \label{tab:comparePUF2}
    \resizebox{1.0\linewidth}{!}{  
\begin{tabular}{c|c|c|c|c|c}
\cline{2-6}
 & \textbf{Type} & \textbf{Name} & \textbf{Weak/Strong} & \textbf{Ref} & \textbf{Comment} 
\\
\toprule
\toprule
\multirow{2}{*}{\begin{tabular}[c]{@{}l@{}}Special \\ fabrication\end{tabular}} & coating & & Weak & \cite{tuyls2006read, skoric2007experimental, skoric2006information} & Smaller number of CRP
\\ \cline{2-2} \cline{4-6}
& Optical & & Strong & \cite{kursawe2009reconfigurable, skoric2007experimental, ruhrmair2013optical} & Difficult to evaluate the uniqeness
\\ \midrule
\multirow{6}{*}{\begin{tabular}[c]{@{}l@{}}Silicon \\ PUF\end{tabular}} & \multirow{2}{*}{\begin{tabular}[c]{@{}l@{}}Delay \\ based\end{tabular}} & Arbiter & Strong & \cite{suh2007physical, morozov2010analysis, fruhashi2011arbiter} & Vulnerable to attacks\\ \cline{3-6}
& & \begin{tabular}[c]{@{}l@{}}Ring  Oscillator\end{tabular} & Weak & \cite{maiti2009improving, shamsoshoara2019ring, yin2009temperature} & Needs large power and space \\ \cline{2-6}
& \multirow{4}{*}{\begin{tabular}[c]{@{}l@{}}Memory \\ based\end{tabular}} & Re-RAM & Strong & \cite{pavan1997flash, cambou2016puf, afghah2018reram} & \begin{tabular}[c]{@{}c@{}}Very sensitive to environmental and voltage \\ fluctuations \end{tabular}
\\ \cline{3-6}
& & Butterfly & Weak & \cite{kumar2008butterfly, ruhrmair2014pufs} & \begin{tabular}[c]{@{}c@{}}unstable adjoining will effect PUF \\ response \end{tabular}
\\ \cline{3-6}
& & SRAM & Weak & \cite{guajardo2007fpga, guajardo2007physical, claes2011comparison,assiri2019key} & Vulnerable to side-channel attacks
\\ \bottomrule \bottomrule
\end{tabular}
}
}
\end{table}

\subsection{
\black{Robustness of PUF Responses}}
\label{subsec:PUFchallenge}
Analog physical parameters which are used to extract fingerprint of a device, are prone to noise and may change due to temperature, supply voltage and other parameters. The digital fingerprint of the PUF may vary due to changes in any of the parameters mentioned before. Differential design techniques are applied in order to mitigate some of the environmental dependencies in a PUF to make it more stable \cite{puftutorial}.

Although differential design techniques may improve reliability, the change in environmental conditions will introduce noise in the PUF output. Noise may cause one or more PUF output bits to be flipped resulting in \textcolor{black}{server not authenticating a valid client resulting in a } false negative \black{authentication. Mechanisms to reduce the \textit{intra hamming distance} of the PUF responses are required to address the various challenges that hinder their perfect reproducibility.} Different error correction coding techniques are being employed to improve the reproducibility of the PUF using ``\textit{Fuzzy Extractor}" schemes elaborated in section \ref{sec:FuzzyExtractors}. \textcolor{black}{These mechanisms should be designed to reduce the noise in a client PUF causing false negative authentication while ensuring that another client device is not mistaken for the authenticated client, i.e., false positive authentication should not increase while reducing false negative authentications. These mechanisms should reduce any false authentication rates while improving the quality of PUF.} Various error correcting codes have been utilized to reduce the intra PUF variation factor and improve the similarity of PUF responses for the same query \cite{puftutorial}. Adding redundant information (parity bit or helper data) will increase the probability of error detection and correction during the challenge response authentication. Linear block codes and 2-D Hamming codes were used earlier in several PUFs. 

\section{\black{PUF-based Security Solutions}}
\label{sec:PUF_interaction_security_attack}

\black{In Section~\ref{sec:Security_Challenges}, we reviewed different possible attacks in IoT networks from various aspects such as applications, architecture, communication, and data planes. Moreover, we also discussed hardware-based attacks and possible solutions in addition to PUFs. In this section, we study the role of PUFs in mitigating different types of attacks.}

\black{We mentioned Controlled PUFS (C-PUFs) in Section~\ref{sec:Security_Challenges} as one possible solution to prevent man-in-the-middle attacks. The authors in \cite{gassend2002controlled} used C-PUFs to increase the robustness and reliability of normal PUFs. 
C-PUFs can be only accessed by an algorithm that is physically linked to the PUF. In this scheme, the C-PUF is achieved using multiple layers. For instance, they placed one random hash function before the PUF to prevent the man-in-the-middle attack by avoiding a \textit{chosen challenge attack} on PUFs. This technique does not allow the attacker to utilize the model-based adversary attack to extract the PUF parameters. Also, an error correcting code is placed after the PUF output to reduce the noisy output measurements which result in more robust responses. Also, the authors located another random hash function after the error correcting code to make the relationship between the response and the physical measurements more complex. The authors proposed the idea of multiple personalities for this controlled-PUF and they showed this C-PUF is robust to prevent man-in-the-middle attacks.}

\begin{figure}[!htbp]
\begin{center}
\includegraphics[width=0.45\columnwidth]{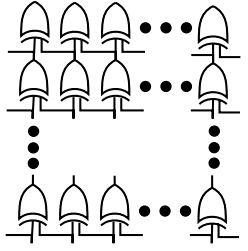}
    \caption{Rectangular PPUF circuit using XOR gates \cite{beckmann2009hardware}.}
    \label{fig:ppuf}
    \end{center}
\end{figure}

\black{The authors in \cite{beckmann2009hardware} proposed a new class of PUF called \textit{Public-PUF (PPUF)} and investigated their role in preventing side channel attacks. They showed that the attacker can reverse engineer this PPUF; however, getting a response or output takes a long time. The authors proposed a rectangular circuit built from XOR gates with a width of $w$ and a height of $h$. The output results from this circuit depends on the gates' delays and even using the simulation it needs a long time to simulate the output. The PPUF needs to reach the steady state point before starting the simulation because of the intermediate gates' delays. As a result, in addition to the simulation time to predict the output, an additional time is required for the circuit to reach a steady state. Figure~\ref{fig:ppuf} shows one simple structure of this PPUF. The authors used these delays to generate the public key in cryptographic methods.}

\black{To evaluate the performance of this PUF, the authors showed that if an attacker utilized a 10 GHz processor and assuming that simulating one number in the PPUF requires $3 \times 10^{12}$ cycles, then simulating a PPUF with $w \approx 10^4$ needs $1.7 \times 10^{29}$ cycles of simulation to extract the secret key. Based on \cite{Howmanyc21_online}, if the attacker uses two billion computers with the power of 10 GHz per computer which is very optimistic, then the total required time is $8.5 \times 10^9$ or 264 years. This approach of PUF is resilient against different physical attacks such as side channels.}

\black{Hardware obfuscation using PUFs is one of the methods that researchers use to avoid the reverse engineering attacks. Using the hardware obfuscation, the designers can hide some part of the design or chips during the fabrication process \cite{koushanfar2012hardware}. The general idea is to keep some modules of the design such that the chip does not work properly during the design. However, it is possible to use an activation function to actuate those modules in a post-fabrication process \cite{zamanzadeh2016higher, rajendran2012security}. Without the activation function or the key, the attacker cannot understand the whole design of the IC, which makes the reverse engineering process less likely. Many authors used PUF to generate a unique key or activation function for each chip in such a way that the attacker cannot use the leaked key for other chips to reverse engineer them \cite{alkabani2007active, wendt2014hardware}. The authors of \cite{khaleghi2018hardware} proposed a new method for the activation key using strong PUFs. The authors proposed that characterizing the strong PUF is almost impossible. The authors showed a scheme where they used a small portion of CRP's space from the strong PUF in the activation phase. As a result, the size of the look-up-table is not costly for the designer; however, the attacker has to deal with the entire space of the strong PUF to extract the key.} 

\black{Hardware Trojans and IP hijacking are common types of security attacks. Many illegal companies are investing on IP hijacking and reverse engineering such as mask theft and illegal overproduction. PUFs can offer effective countermeasures against these kind of attacks. The authors of \cite{dupuis2014novel} proposed logic encryption for ICs. Encryption techniques can be categorized in two types, first) Some logic gates will be added to the main circuit to hide the true logic and behavior of the circuit, second) A set of specific sequences of inputs are required to reach a valid state of functionality. In this case, the state transition will be modified such that the attacker cannot predict the behavior. The authors used a combinational method to minimize the probability of \textit{rare values}. \textit{Rare values} are those true low controllability signals and conditions that can trigger the inserted Trojan. To design the encryption algorithm, the authors used PUF with additional logic gates to generate a random unique key for each IC. The authors claimed that using this PUF and encryption method, the attacker cannot exploit those \textit{rare values} to initiate the Trojan.}

\section{Fuzzy Extractors}
\label{sec:FuzzyExtractors}
\textcolor{black}{Fuzzy extractors (FE) are mechanisms which help to extract original registered responses from noisy responses. In \cite{dodis2004fuzzy}, the authors proposed the FE in order to correct noisy biometric responses. To correct the noisy response, the initial response of the client needs to be registered. During the registration process,} from an initial input $w$,  a uniformly random string $R$ and non-secret string $P$ (helper data) are extracted using a fuzzy extractor. This mechanism allows the string $R$ to be used as a key and reproduced exactly with the help of $P$, even though the input changes to some $w'$ but remains close to $w$. These mechanisms are said to be \textit{information-theoretic secure}, i.e., a crypto-system whose security is derived only from information theory, where the adversary cannot break the encryption due to insufficient information, thereby allowing them to be used in cryptography.
\begin{figure}[!htbp]
\begin{center}
\includegraphics[width=0.9\columnwidth]{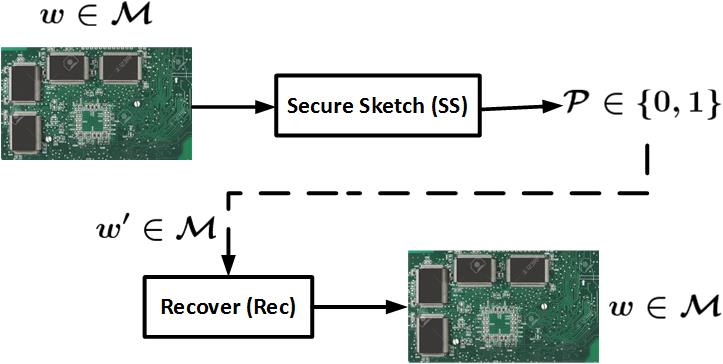}
    \caption{Secure sketch}
    \label{fig: fuzzyExtractor_a}
    \end{center}
\end{figure}

FEs are constructed using \textit{Secure Sketch} (SS), which is a combination of two procedures ``sketch'' and ``recover'' that allow precise reconstruction of the initial input from noisy input by making use of some helper data $P$.

\textit{Helper Data} $P$ is extracted from initial input $w$, which can be made publicly available in the ``sketch" phase. This output $P$ will be used in the ``recover" phase along with noisy input $w'$ to recover $w$. This method is secure because the publicly available \textit{Helper Data} reveals little to no information about $w$. Figure \ref{fig: fuzzyExtractor_a} describes a secure sketch. 

\begin{figure}[!htbp]
\begin{center}
	\includegraphics[width=\columnwidth]{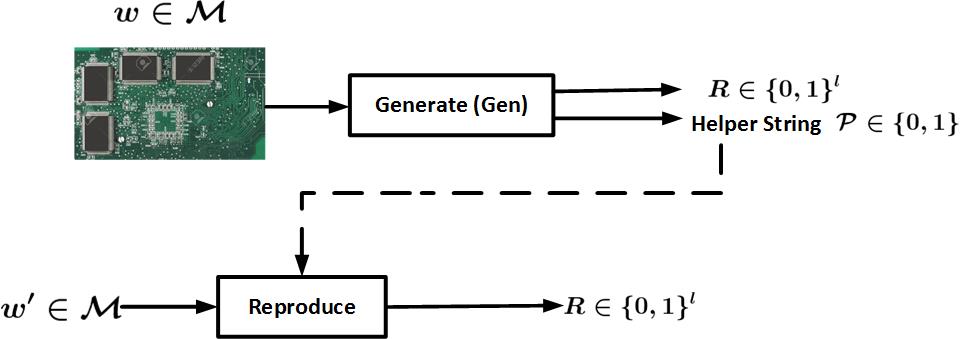}
    \caption{Block diagram of a Fuzzy Extractor}
    \label{fig: fuzzyExtractor_b}
    \end{center}
\end{figure}

\begin{figure}[!htbp]
	\includegraphics[width=\columnwidth]{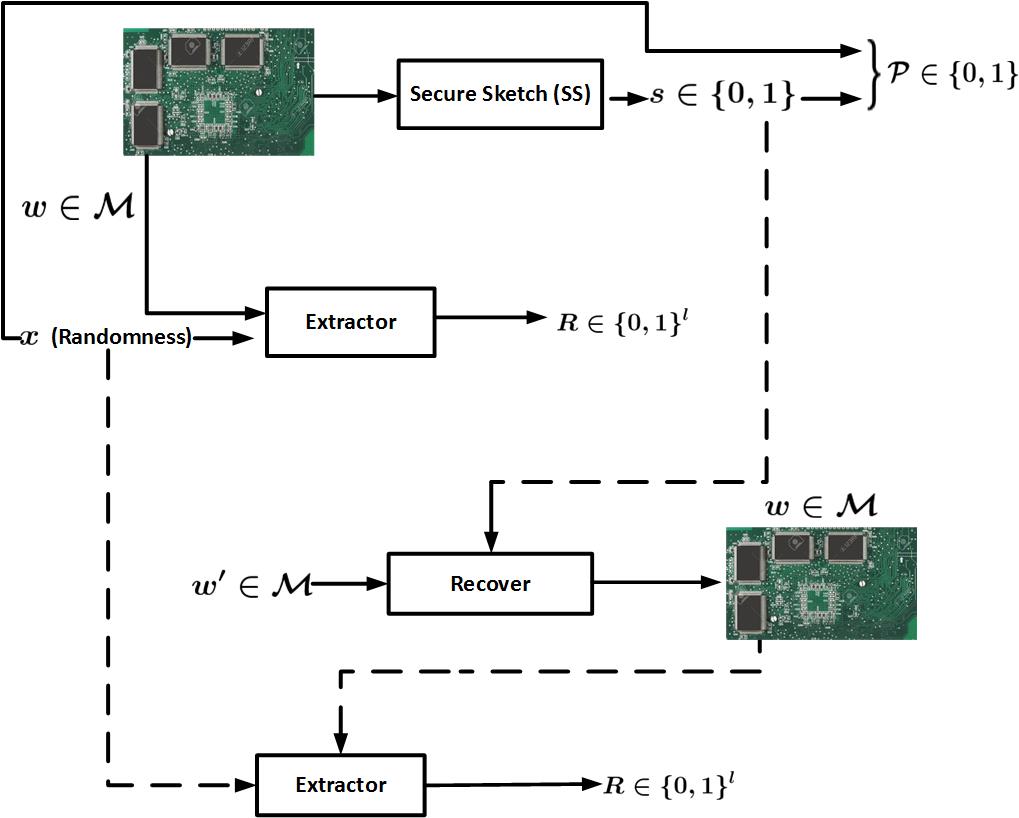} 
    \caption{ Construction of a fuzzy extractor using secure sketch extractor}
    \label{fuzzyExtractorConstruction}
\end{figure}

FE is defined with a combination of a pair of procedures called ``generate'' and ``reproduce". In the ``generate" phase, the fuzzy extractor uses the ``sketch'' phase of the SS where \textit{Helper data}, $P$ and \textit{Key}, $R$ are extracted from the given input $w$. The ``reproduce" phase uses the ``recover'' phase of the secure sketch which makes use of the \textit{Helper data} to reproduce the \black{initial} 
input $w$ from a noisy input $w'$ along with the random extractor used in the ``sketch" phase, to extract the randomness from the recovered $w$. The ability to recover $w$ from $w'$ is \black{highly dependent on the technique used to correct the data. In this case, based on the initial FE proposed by Dodis et al., in \cite{dodis2004fuzzy}, the authors utilized the concepts of correcting the received communication message by using error correction codes. All the other FEs proposed in the literature also followed the same trend, utilizing different types of error correction codes to correct noisy input data. } 
\black{These error correcting codes are used} in the ``sketch'' phase of the FE. If the \black{hamming} 
distance between the noisy input $w'$ and input $w$ is too large, it may not be possible to recover $w$ from $w'$. Figure \ref{fuzzyExtractorConstruction} shows the construction of a FE using a secure sketch. 
As Secure Sketches have an error tolerance capability, error correcting codes are utilized in their construction. The error correcting code $C$ is used to correct errors in $w'$ by shifting the codeword, although $w'$ may not be in $C$. Two different constructions are used for secure sketch are presented \cite{dodis2004fuzzy}: 
\begin{itemize}
\item{\textit{``Code-Offset Construction: For input $w$, a uniformly random codeword $c$ is selected from $C$, and SS(w) to be the shift needed to get from $c$ to $w$: $SS(w)=w-c$. To compute $Rec(w',s)$,  the shift $s$ is subtacted from $w'$ to get $c'= w'-s$: $c'$ is decoded to retrieve $c$ and $w$ is computed by shifting back to get $w=c+s$. When code $C$ is linear, the information in s is essentially the syndrome of w."}\cite{dodis2004fuzzy}}
\item{\textit{``Syndrome Construction: The sketch SS(w) computes $s= \textit{syn}(w)$, where \textit{syn} is the syndrome. To recover the key, a unique vector $e$ is chosen such that $syn(e)= syn(w')-s$ and output $w=w'-e$."}}
\end{itemize}

Bose-Chaudhuri-Hocquenghem (BCH) codes which are one of the cyclic error correcting codes, have been implemented in fuzzy extractors, where 192 syndrome bits out of 255 bits were exposed as helper data offered a stability of 88\% to a PUF correcting 30 errors in a noisy PUF \cite{suh2005}. High performance error correction coding schemes such as Convolution Coding and low density parity check (LDPC) are applicable to PUF error correction but are highly complex to implement. It is practical to use good error correction codes which consider maximum likelihood estimation, implementation complexity and secret key leakage to improve the performance of the PUF. 

\section{Generation of cryptographic keys from PUFs using Fuzzy Extractors} 
\label{sec:KeyGenFuzzyExtractors}
Secret key generation using PUFs will allow users to produce a key from their own devices which need not be stored in the device. Using PUFs to produce keys will make the device unclonable and hence less susceptible to hacking. Additionally, the use of PUFs eliminates the security issues related to key storage and distribution. Different PUFs have been used in the past to generate reliable and reproducible cryptographic keys using FE.

Different methods to generate reproducible keys using PUFs have been proposed in the literature \cite{suh2007physical, kang2014performance,kursawe2009reconfigurable,ziola2014authentication,vskoric2005robust, lim2005extracting}. 
The key generation scheme proposed in \cite{kang2014performance} uses BCH codes and random number generators for the construction of fuzzy extractors. The main goal of BCH codes is to help reconstruct the PUF estimate from noisy PUF data in order to be used in the ``secure sketch" phase of the FE. 

The error correction capability of the fuzzy extractor is dependant on the strength of the decoder to recover the message. This strength generally enhances with increasing the complexity of the system. If a system with high complexity is being utilized in a fuzzy extractor, there is a good chance that the system is not only correcting all the errors, but might over correct the message thereby introducing more errors into the message. This scheme may be suitable for authentication where a certain error margin can be tolerated. In key generation schemes, the key which is used for encrypting needs to be reproduced with a zero error margin in order to decrypt the encrypted data. 

\black{Many papers in the literature have proposed different FE architectures. In \cite{maes2012pufky}, a practical and modular design for key generation using a PUF was proposed which was evaluated on a set of FPGA devices on a ring oscillator PUF. This fuzzy extractor was able to produce a secure 128-bit key with a failure rate $< 10^{-9}$ in 5.62 ms. In this research, the author used a repetition based BCH code, where the syndrome generation and error decoding are fully combinatorial. 
}

\black{In \cite{bosch2008efficient}, an implementation of a fuzzy extractor on a FPGA was done in order to measure the efficiency in terms of the required hardware resources. The authors proposed decoder architectures for Reed Muller and Golay codes. In \cite{maes2009low}, a helper data algorithm using soft decision information was proposed. The authors argued that this technique could lower the entropy loss of a helper data algorithm. The authors chose to use soft decision maximum likelihood decoding and generalized multiple concatenated codes as other soft decision decoders such as belief propagation for LDPC and Viterbi algorithm for convolutional codes that require very long data streams.} 

\black{In \cite{puchinger2015error}, a helper data algorithm based on generalized concatenated, Reed Muller and Reed Solomon codes was proposed, where a simplex inner code (16,5,8) was selected with a RM outer code of 128 bits to obtain a codeword of length $128\times{16}$ = 2048.}
\begin{table}[ht]
\caption{Comparison of failure probability and flipping probability of different fuzzy extractor schemes proposed in the literature.\cite{korenda2018secret}}
	\label{fuzzyTable}
\resizebox{\columnwidth}{!}{%
		\begin{tabular}{|c|c|c|c|c|}
			\hline
		
			\multicolumn{1}{|m{5cm}|}{Fuzzy extractor construction}  & Key length    & Helper data bits           & Failure probability  & Flipping probability          \\ 
			\hline
				\multicolumn{1}{|m{5cm}|}{BCH repetition code	\cite{maes2012pufky}}	    &128  & 2052   & $10^{-9}$ & 13\%      \\ \hline
			\multicolumn{1}{|m{5cm}|}{Reed Muller generalized multiple concatenated coding	\cite{maes2009low}}	  &128    & 13952   & $10^{-6}$ & 15\%  \\ 
			\hline
			
			\multicolumn{1}{|m{5cm}|}{Generalized concatenated (GC) Reed Muller\cite{puchinger2015error} }		& 2048     & 2048   & $5.37. 10^{-10}$ & 14\%     \\
			 \hline
			\multicolumn{1}{|m{5cm}|}{GC Reed Solomon\cite{puchinger2015error}}                         &1024 & 1024 &  $3.47. 10^{-10}$ & 14\% \\
			 \hline
			\multicolumn{1}{|m{5cm}|}{Polar codes with SC \cite{chen2017high}}       &128                 & 896             &  $ 10^{-6}$     &15\%       \\ 
			\hline
			\multicolumn{1}{|m{5cm}|}{Polar codes with Hash-Aided SCL decoder \cite{chen2017high}}       &128                  & 896             &  $ 10^{-9}$     &15\%       \\ 
			\hline
 			\multicolumn{1}{|m{5cm}|}{{ Serially concatenated BCH-Polar codes with SC decoder \cite{korenda2018secret}}  }   & {250 }                         & {\bf 262}          & $ {\bf 10^{-8
			}}$  & 15\%        \\ 
            \hline
 		\multicolumn{1}{|m{5cm}|}{	{ Serially concatenated BCH-Polar codes with belief propagation decoder \cite{korenda2018secret}}  }    & {250 }                         & { 262}          & $ { 10^{-10
			}}$  & 15\%        \\ \hline
		\end{tabular}%
        }
\end{table}

In a recent research work proposed in \cite{chen2017high}, a FE structure based on Polar codes for SRAM PUFs was proposed. This work utilized complex Hash-Aided SC decoder to ensure that the key was reproducible. Using this methods the authors in \cite{chen2017high} were able to reproduce their key using 896 helper bits, with a failure probability of $10^{-9}$ for a key length of 128 bits. In \cite{Korenda_SECON}, we used a custom-built Arduino shield to read the fingerprint of SRAM PUF devices. We later compared the efficiency of different fuzzy extractor schemes to generate reliable and reproducible keys while estimating the FAR and FRR. 
The FE techniques proposed in the literature utilize long Helper data bits and complex decoder structures to extract the low failure probability rates. There is a high probability in increasing the FAR and FRR of the PUF based Keys when using high complex- sophisticated decoders. \textcolor{black}{These decoders were} designed to extract the message from very noisy channels, which may not be the best solution to PUF based keys.  \textcolor{black}{Using decoders which were designed to extract messages from very noisy channels may cause the fuzzy extractor to estimate the PUF response in an incorrect way leading to an increase in the false authentication rate.} Further research to understand the average error rate of each PUF and using mechanisms appropriate for the PUF will help mitigate such situations. Table \ref{fuzzyTable}  presents a comparison of the results obtained by these different architectures. 

\textcolor{black}{Fuzzy extractors may be used to extract a PUF response using a noisy PUF response, but many research efforts have focused on attacking a PUF using a fuzzy extractor as they produce the helper data. This is due to the fact, that while the publicly available \textit{helper data} is used to recover the noisy responses in the client side, it can also be a primary target for attackers to extract the response from the PUF. These extractions can further lead to characterizing the entire device if enough responses can be extracted from a single PUF. Some of the attacks on Fuzzy extractors using ``helper data manipulation" are discussed in section \ref{sec:FEAttacks}.}

\black{In general, all the Key generation protocols proposed strive to achieve a probability of error of $10^{-6}$ to $10^{-10}$. The tolerable error rate for a PUF-based key generation scheme is highly dependent on the application the device utilizes. When used for a sensitive application, a very low probability of error in the range of $10^{-10}$ is desired which is possible at a cost of increased complexity. Therefore, there is a trade-off between the desired complexity and the probability of error.}

\black{The concept of on-demand key generation has also gained huge popularity in recent years. Theoretically, the number of attacks possible on a PUF can also be reduced when a different key is utilized each time a message is transmitted. There are many factors that depend on the generation of a new key. Some of these factors include the entropy space of the PUF, time delay to read the PUF response, time delay caused in the generation of a key using a FE, and so on. A strong PUF has a high entropy and will be able to generate multiple keys for each message while a weak PUF has to repeat the same keys over a period of time. 
In a SRAM PUF, there is an additional delay cost because one has to wait for the PUF to turn ON/OFF in order to read the PUF response. This cost is also a key factor in generating the on-demand-keys. The fuzzy extractor utilized will also have to extract a new key each time a new PUF response is chosen and store it in the server for decryption which also adds to the time delay. If the application can tolerate the added delay, and the entropy of the PUF allows it to generate on-demand-keys, then this idea will be popular in the cryptographic community. This is an open challenge in this area, where researchers are constantly trying to increase the entropy of the PUF while reducing the delay caused by on-demand key generation. Depending on the application, a trade-off should be established between the delay caused due to current hardware technology and the need for a new key each time it is required.}

\section{\black{Attacks on Fuzzy Extractors via Helper Data Manipulation}}\label{sec:FEAttacks}
\black{
In \cite{merli2011side}, the authors proposed a method to extract the cryptographic keys from PUF using helper data manipulation attacks by tracking the power or electromagnetic measurements of the power consumed when manipulated versions of helper are used to generate keys. The value of power is extracted from every trace and the bit pattern corresponding to minimum power is considered to be the correct codeword. 
}

\black{As ring oscillator (RO) PUFs are easily deployable on FPGAs. ROs have been gaining a lot of attention in many research works for some time.  In \cite{delvaux2014key}, the authors had demonstrated manipulation attacks on temperature aware cooperative RO PUFs \cite{yin2009temperature}, sequential pairing algorithm \cite{yin2010lisa}, group-based RO PUFs \cite{yin2013design}, all entropy distiller constructions \cite{yin2013improving}. The attacks proposed in this paper statistically estimate the failure rate when the Helper data is manipulated to extract the key. In temperature aware cooperative RO PUFs, errors were injected into the Helper data which comprises cooperating pairs, while calculating the failure rate. In a similar way, steep polynomials were injected into the entropy distiller, Helper data in Group RO PUF, to overshadow the random frequency variations. In  Sequentially paired RO PUFs, a sequential pairing algorithm is used to select disjunct pairs. In order to attack this RO PUF, response bits are matched with other responses and the failure rate is calculated to understand the secret key.}

\black{Different constructions of error correction codes used in FE are discussed in \cite{delvaux2014helper}. Some of their findings include new threats using Helper data algorithms  in terms of leakage and manipulation. The authors also derive a way to accurately calculate the leakage of repetition codes in case of bias. The leakage caused due to soft decision coding is underestimated according to the authors. The authors also propose a "divide and conquer manipulation attack" for parallel, concatenated and soft decision codes. }
     
      

\textcolor{black}{In \cite{delvaux2014attacking}, the authors discuss using the Helper data manipulation attack on Pattern Matching Key Generators (PMKG) which was proposed in \cite{paral2011reliable}. PMKGs utilize sub-strings of a long PUF response stream called \textit{patterns}. In these mechanisms, the indices of the sub string are stored as the \textit{Key} while the patterns are stored as public \textit{helper data}. The key is reconstructed using a matching procedure when a regenerated response is inserted across the patterns. In \cite{delvaux2014attacking}, the full bitstreams of a PUF and their secret indices were extracted by manipulating the Helper data and statistically observing the failure rate of a PMKG. The attack was further demonstrated on a 4-XOR arbiter PUF.  
}

\section{Conclusion and Future Discussion}
\label{sec:conclusion}
This survey paper reviews different security challenges in IoT networks and devices. Different domains such as data, communication, architecture, and application are considered for the security taxonomy in IoTs. Software attacks are discussed based on these aforementioned domains. 
Then, we reviewed the semiconductor manufacturing process chain which consists of different tasks to accomplish the hardware design for ICs. This chain brings various vulnerabilities in different points which can be considered as hardware attacks. PUFs are considered as one of the potential solutions to counter the hardware based attacks. Noting the challenges of using PUFs due to the variations between different responses. We surveyed various methods to extract keys from noisy PUF responses  using fuzzy extractors schemes. The importance and need for using  the fuzzy extractors schemes for generating cryptographic keys from PUFs is discussed. A brief discussion on the probability of FE schemes to over correct the Noisy PUF responses and leading to false authentication rate or false rejection rate is discussed.

The future directions of using PUFs need to focus not only on extracting reproducible schemes from Noisy PUFs but also to understand the behaviour of each PUF for different environmental and physical conditions. This challenge required designing schemes that will tailor themselves to the expected error in each device, thereby not allowing under correction or over correction of the PUF responses; thereby reducing the False Rejection Rate and False Authentication Rate of the PUF.

\section{Acknowledgments}
The authors would like to thank the anonymous reviewers for their valuable comments which helped us improve the organization, content, and quality of this paper.
This material is based upon the work supported by the National Science Foundation under Grant No. 1827753.


\bibliography{mybibfile}

\begin{thebibliography}{100}
\expandafter\ifx\csname url\endcsname\relax
  \def\url#1{\texttt{#1}}\fi
\expandafter\ifx\csname urlprefix\endcsname\relax\def\urlprefix{URL }\fi
\expandafter\ifx\csname href\endcsname\relax
  \def\href#1#2{#2} \def\path#1{#1}\fi

\bibitem{bhayani2016internet}
M.~Bhayani, M.~Patel, C.~Bhatt, {Internet of {Things} ({IoT}): In a way of
  Smart World}, in: Proceedings of the international congress on information
  and communication technology, Springer, 2016, pp. 343--350.

\bibitem{galleso2016samsung}
M.~Galleso, {Samsung Gear S3 Classic and Frontier: An Easy Guide to Best
  Features}, Lulu Press, Inc, 2016.

\bibitem{lu2011application}
D.~Lu, T.~Liu, {The {A}pplication of {IoT} in {M}edical {S}ystems}, in: IT in
  Medicine and Education (ITME), 2011 International Symposium on, Vol.~1, IEEE,
  2011, pp. 272--275.

\bibitem{kloti2013openflow}
R.~Kloti, V.~Kotronis, P.~Smith, {Openflow: A Security Analysis}, in: {Network
  Protocols (ICNP), 2013 21st IEEE International Conference on}, IEEE, 2013,
  pp. 1--6.

\bibitem{zhang2014iot}
Z.-K. Zhang, M.~C.~Y. Cho, C.-W. Wang, C.-W. Hsu, C.-K. Chen, S.~Shieh, {{IoT}
  Security: Ongoing Challenges and Research Opportunities}, in:
  {Service-Oriented Computing and Applications (SOCA), 2014 IEEE 7th
  International Conference on}, IEEE, 2014, pp. 230--234.

\bibitem{fuhong2014cooperative}
L.~Fuhong, L.~Qian, Z.~Xianwei, C.~Yueyun, H.~Daochao, {Cooperative
  Differential Game for Model Energy-bandwidth Efficiency Tradeoff in the
  Internet of Things}, China Communications 11~(1) (2014) 92--102.

\bibitem{valehi2017graph}
A.~Valehi, A.~Razi, B.~Cambou, W.~Yu, M.~Kozicki, {A Graph Matching Algorithm
  for User Authentication in data Networks using Image-based Physical
  Unclonable Functions}, in: 2017 Computing Conference, IEEE, 2017, pp.
  863--870.

\bibitem{valehi2017maximizing}
A.~Valehi, A.~Razi, {Maximizing Energy Efficiency of Cognitive Wireless Sensor
  Networks with Constrained Age of Information}, IEEE Transactions on Cognitive
  Communications and Networking 3~(4) (2017) 643--654.

\bibitem{framling2014universal}
K.~Fr{\"a}mling, S.~Kubler, A.~Buda, {Universal Messaging Standards for the IoT
  From a Lifecycle Management Perspective.}, IEEE Internet of {Things} Journal
  1~(4) (2014) 319--327.

\bibitem{shamsoshoara2015enhanced}
A.~Shamsoshoara, Y.~Darmani, {Enhanced Multi-route ad hoc On-demand Distance
  Vector Routing}, in: Electrical Engineering (ICEE), 2015 23rd Iranian
  Conference on, IEEE, 2015, pp. 578--583.

\bibitem{HanUAVMobility}
H.~{Peng}, A.~{Razi}, F.~{Afghah}, J.~{Ashdown}, A unified framework for joint
  mobility prediction and object profiling of drones in uav networks, Journal
  of Communications and Networks 20~(5) (2018) 434--442.

\bibitem{sheng2013survey}
Z.~Sheng, S.~Yang, Y.~Yu, A.~Vasilakos, J.~Mccann, K.~Leung, {A Survey on the
  {IETF} Protocol Suite for the {Internet of Things}: Standards, Challenges,
  and Opportunities}, IEEE Wireless Communications 20~(6) (2013) 91--98.

\bibitem{shamsoshoara2019distributed}
A.~Shamsoshoara, M.~Khaledi, F.~Afghah, A.~Razi, J.~Ashdown, {Distributed
  Cooperative Spectrum Sharing in UAV networks using Multi-agent Reinforcement
  Learning}, in: 2019 16th IEEE Annual Consumer Communications \& Networking
  Conference (CCNC), IEEE, 2019, pp. 1--6.

\bibitem{shamsoshoara2019solution}
A.~Shamsoshoara, M.~Khaledi, F.~Afghah, A.~Razi, J.~Ashdown, K.~Turck, {A
  Solution for Dynamic Spectrum Management in Mission-Critical UAV Networks},
  arXiv preprint arXiv:1904.07380.

\bibitem{8406970}
F.~Afghah, A.~Shamsoshoara, L.~Njilla, C.~Kamhoua, {A reputation-based
  {Stackelberg} game model to enhance secrecy rate in spectrum leasing to
  selfish {IoT} devices}, in: {IEEE INFOCOM 2018 - IEEE Conference on Computer
  Communications Workshops (INFOCOM WKSHPS)}, 2018, pp. 312--317.
\newblock \href {http://dx.doi.org/10.1109/INFCOMW.2018.8406970}
  {\path{doi:10.1109/INFCOMW.2018.8406970}}.

\bibitem{afghah2020cooperative}
F.~Afghah, A.~Shamsoshoara, L.~L. Njilla, C.~A. Kamhoua, Cooperative spectrum
  sharing and trust management in iot networks, Modeling and Design of Secure
  Internet of Things (2020) 79--109.

\bibitem{kamhouamodeling}
C.~A. Kamhoua, L.~L. Njilla, A.~Kott, Modeling and design of secure internet of
  things, Wiley Online Library.

\bibitem{kamhoua2020modeling}
C.~Kamhoua, L.~Njilla, A.~Kott, S.~Shetty,
  \href{https://books.google.com/books?id=-jXrDwAAQBAJ}{Modeling and Design of
  Secure Internet of Things}, Wiley, 2020.
\newline\urlprefix\url{https://books.google.com/books?id=-jXrDwAAQBAJ}

\bibitem{AsIoTatt94:online}
A.~D. Rayome, {As IoT attacks increase 600\% in one year, businesses need to up
  their security - TechRepublic},
  \url{https://www.techrepublic.com/article/as-iot-attacks-increase\\-600-in-one-year-businesses-need-to-up-their-security/},
  (Accessed on 09/26/2018) (March 2018).

\bibitem{IoTnumb86:online}
S.~2020, {IoT: number of connected devices worldwide 2012-2025 | Statista},
  \url{https://www.statista.com/statistics/471264/iot-number-of-\\connected-devices-worldwide/},
  (Accessed on 04/27/2020).

\bibitem{doffman2019}
Z.~Doffman, {Cyberattacks On IOT Devices Surge 300\% In 2019, ‘Measured In
  Billions’, Report Claims},
  \url{https://www.forbes.com/sites/zakdoffman/2019/09/14/dangerous-\\cyberattacks-on-iot-devices-up-300-in-2019-now-rampant-report\\-claims/\#3e0d17195892},
  (Accessed on 04/28/2020) (Sep 2019).

\bibitem{WannaCry56:online}
B.~Vigliarolo, {WannaCry: A cheat Sheet for Professionals - TechRepublic},
  \url{https://www.techrepublic.com/article/wannacry-the-smart-\\persons-guide/},
  (Accessed on 09/26/2018) (june 2017).

\bibitem{o2012ransomware}
G.~O'Gorman, G.~McDonald, {Ransomware: A Growing Menace}, Symantec Corporation,
  2012.

\bibitem{brewer2016ransomware}
R.~Brewer, {Ransomware Attacks: Detection, Prevention and Cure}, Network
  Security 2016~(9) (2016) 5--9.

\bibitem{Petyaran66:online}
N.~Heath, {Petya Ransomware: Where it Comes from and How to Protect yourself},
  \url{https://www.techrepublic.com/article/petya-ransomware-where\\-it-comes-from-and-how-to-protect-yourself/},
  (Accessed on 09/26/2018) (June 2017).

\bibitem{sapienza2017early}
A.~Sapienza, A.~Bessi, S.~Damodaran, P.~Shakarian, K.~Lerman, E.~Ferrara,
  {Early Warnings of Cyber Threats in Online Discussions}, in: {Data Mining
  Workshops (ICDMW), 2017 IEEE International Conference on}, IEEE, 2017, pp.
  667--674.

\bibitem{yaqoob2017rise}
I.~Yaqoob, E.~Ahmed, M.~H. ur~Rehman, A.~I.~A. Ahmed, M.~A. Al-garadi,
  M.~Imran, M.~Guizani, {The Rise of Ransomware and Emerging Security
  Challenges in the Internet of Things}, Computer Networks 129 (2017) 444--458.

\bibitem{DynAnaly42:online}
S.~Hilton, {Dyn Analysis Summary Of Friday October 21 Attack | Dyn Blog},
  \url{https://dyn.com/blog/dyn-analysis-summary-of-friday-october-\\21-attack/},
  (Accessed on 04/01/2020) (October 2016).

\bibitem{dynCyber2016}
\href{https://en.wikipedia.org/wiki/2016\_Dyn\_cyberattack}{{Dyn Cyberattack}}
  (Jul 2018).
\newline\urlprefix\url{https://en.wikipedia.org/wiki/2016\_Dyn\_cyberattack}

\bibitem{ACC}
\href{https://www.acc.org/about-acc/press-releases/2018/02/20\\/13/57/can-your-cardiac-device-be-hacked}{{Can
  Your Cardiac Device Be Hacked?}} (February 2018).
\newline\urlprefix\url{https://www.acc.org/about-acc/press-releases/2018/02/20\\/13/57/can-your-cardiac-device-be-hacked}

\bibitem{samie2016computation}
F.~Samie, V.~Tsoutsouras, L.~Bauer, S.~Xydis, D.~Soudris, J.~Henkel,
  {Computation Offloading and Resource Allocation for Low-power IoT Edge
  Devices}, in: Internet of Things (WF-IoT), 2016 IEEE 3rd World Forum on,
  IEEE, 2016, pp. 7--12.

\bibitem{dodis2004fuzzy}
Y.~Dodis, L.~Reyzin, A.~Smith, {Fuzzy Extractors: How to Generate Strong Keys
  from Biometrics and Other Noisy Data}, in: {International conference on the
  theory and applications of cryptographic techniques}, Springer, 2004, pp.
  523--540.

\bibitem{shamsoshoara2019overview}
A.~Shamsoshoara, {Overview of Blakley's Secret Sharing Scheme}, arXiv preprint
  arXiv:1901.02802.

\bibitem{chou201711}
S.-Y. Chou, Y.-S. Chen, J.-H. Chang, Y.-D. Chih, T.-Y.~J. Chang, {11.3 A 10nm
  32Kb Low-voltage Logic-compatible Anti-fuse One-time-programmable Memory with
  Anti-tampering Sensing Scheme}, in: {Solid-State Circuits Conference (ISSCC),
  2017 IEEE International}, IEEE, 2017, pp. 200--201.

\bibitem{joye2008white}
M.~Joye, {On White-box Cryptography}, Security of Information and Networks
  (2008) 7--12.

\bibitem{kinney2006trusted}
S.~L. Kinney, {Trusted Platform Module Basics: using TPM in Embedded Systems},
  Elsevier, 2006.

\bibitem{gerjuoy2005shor}
E.~Gerjuoy, {Shor’s Factoring Algorithm and Modern Cryptography: An
  Illustration of the Capabilities Inherent in Quantum Computers}, American
  journal of physics 73~(6) (2005) 521--540.

\bibitem{chen2016report}
L.~Chen, L.~Chen, S.~Jordan, Y.-K. Liu, D.~Moody, R.~Peralta, R.~Perlner,
  D.~Smith-Tone, {Report on Post-quantum Cryptography}, US Department of
  Commerce, National Institute of Standards and Technology, 2016.

\bibitem{amy2016estimating}
M.~Amy, O.~Di~Matteo, V.~Gheorghiu, M.~Mosca, A.~Parent, J.~Schanck,
  {Estimating the Cost of Generic Quantum pre-image Attacks on SHA-2 and
  SHA-3}, in: International Conference on Selected Areas in Cryptography,
  Springer, 2016, pp. 317--337.

\bibitem{mitchell2019impact}
C.~J. Mitchell, {The Impact of Quantum Computing on Real-world Security: A 5G
  Case Study}, arXiv preprint arXiv:1911.07583.

\bibitem{mavroeidis2018impact}
V.~Mavroeidis, K.~Vishi, M.~D. Zych, A.~J{\o}sang, {The Impact of Quantum
  Computing on Present Cryptography}, arXiv preprint arXiv:1804.00200.

\bibitem{barker2011transitions}
E.~Barker, A.~Roginsky, {Transitions: Recommendation for Transitioning the use
  of Cryptographic Algorithms and Key Lengths}, NIST Special Publication 800
  (2011) 131A.

\bibitem{gassend2002silicon}
B.~Gassend, D.~Clarke, M.~Van~Dijk, S.~Devadas, {Silicon Physical Random
  Functions}, in: {Proceedings of the 9th ACM conference on Computer and
  communications security}, ACM, 2002, pp. 148--160.

\bibitem{chatterjee2018building}
U.~Chatterjee, V.~Govindan, R.~Sadhukhan, D.~Mukhopadhyay, R.~S. Chakraborty,
  D.~Mahata, M.~M. Prabhu, {Building PUF based Authentication and Key Exchange
  Protocol for IoT without Explicit CRPs in Verifier Database}, IEEE
  Transactions on Dependable and Secure Computing.

\bibitem{mcgrath2019puf}
T.~McGrath, I.~E. Bagci, Z.~M. Wang, U.~Roedig, R.~J. Young, {A puf taxonomy},
  Applied Physics Reviews 6~(1) (2019) 011303.

\bibitem{ehret2019security}
A.~{Ehret}, K.~{Gettings}, B.~R. {Jordan}, M.~A. {Kinsy}, {A Survey on Hardware
  Security Techniques Targeting Low-Power SoC Designs}, in: 2019 IEEE High
  Performance Extreme Computing Conference (HPEC), 2019, pp. 1--8.

\bibitem{delvaux2015survey}
J.~Delvaux, R.~Peeters, D.~Gu, I.~Verbauwhede, {A survey on lightweight entity
  authentication with strong PUFs}, ACM Computing Surveys (CSUR) 48~(2) (2015)
  1--42.

\bibitem{babaei2019physical}
A.~Babaei, G.~Schiele, {Physical Unclonable Functions in the Internet of
  Things: State of the Art and Open Challenges}, Sensors 19~(14) (2019) 3208.

\bibitem{burg2018wireless}
A.~Burg, A.~Chattopadhyay, K.-Y. Lam, {Wireless Communication and Security
  Issues for Cyber--Physical Systems and the Internet-of-Things}, Proceedings
  of the IEEE 106~(1) (2018) 38--60.

\bibitem{wolf2018safety}
M.~Wolf, D.~Serpanos, {Safety and Security in Cyber-physical Systems and
  Internet-of-Things Systems}, Proceedings of the IEEE 106~(1) (2018) 9--20.

\bibitem{lin2017survey}
J.~Lin, W.~Yu, N.~Zhang, X.~Yang, H.~Zhang, W.~Zhao, {A survey on Internet of
  Things: Architecture, Enabling Technologies, Security and Privacy, and
  Applications}, IEEE Internet of Things Journal 4~(5) (2017) 1125--1142.

\bibitem{arshad2018recent}
S.~Arshad, M.~A. Azam, M.~H. Rehmani, J.~Loo, {Recent Advances in
  Information-centric Networking based Internet of Things (ICN-IoT)}, IEEE
  Internet of Things Journal.

\bibitem{granjal2015security}
J.~Granjal, E.~Monteiro, J.~S. Silva, {Security for the Internet of Things: a
  Survey of Existing Protocols and Open Research Issues}, IEEE Communications
  Surveys \& Tutorials 17~(3) (2015) 1294--1312.

\bibitem{chen2011survey}
H.~Chen, Y.~Chen, D.~H. Summerville, {A Survey on the Application of FPGAs for
  Network Infrastructure Security}, IEEE Communications Surveys \& Tutorials
  13~(4) (2011) 541--561.

\bibitem{sfyrakis2020survey}
I.~Sfyrakis, T.~Gross, A survey on hardware approaches for remote attestation
  in network infrastructures, arXiv preprint arXiv:2005.12453.

\bibitem{chowdhury2020physical}
S.~Chowdhury, A.~Covic, R.~Y. Acharya, S.~Dupee, F.~Ganji, D.~Forte, Physical
  security in the post-quantum era: A survey on side-channel analysis, random
  number generators, and physically unclonable functions, arXiv preprint
  arXiv:2005.04344.

\bibitem{mendez2017internet}
D.~M. Mendez, I.~Papapanagiotou, B.~Yang, {Internet of Things: Survey on
  security and privacy}, arXiv preprint arXiv:1707.01879.

\bibitem{alaba2017internet}
F.~A. Alaba, M.~Othman, I.~A.~T. Hashem, F.~Alotaibi, {Internet of Things
  security: A survey}, Journal of Network and Computer Applications 88 (2017)
  10--28.

\bibitem{zhao2013survey}
K.~Zhao, L.~Ge, {A Survey on the Internet of Things Security}, in:
  {Computational Intelligence and Security (CIS), 2013 9th International
  Conference on}, IEEE, 2013, pp. 663--667.

\bibitem{el2016internet}
O.~El~Mouaatamid, M.~Lahmer, M.~Belkasmi, {Internet of Things Security: Layered
  classification of Attacks and Possible Countermeasures}, electronic journal
  of information technology 9~(9).

\bibitem{botta2014integration}
A.~Botta, W.~De~Donato, V.~Persico, A.~Pescap{\'e}, {On the Integration of
  Cloud Computing and Internet of Things}, in: {2014 International Conference
  on Future Internet of Things and Cloud}, IEEE, 2014, pp. 23--30.

\bibitem{hashem2016role}
I.~A.~T. Hashem, V.~Chang, N.~B. Anuar, K.~Adewole, I.~Yaqoob, A.~Gani,
  E.~Ahmed, H.~Chiroma, {The Role of Big Data in Smart City}, International
  Journal of Information Management 36~(5) (2016) 748--758.

\bibitem{keshavarz2018towards}
M.~Keshavarz, M.~Anwar, Towards improving privacy control for smart homes: A
  privacy decision framework, in: 2018 16th Annual Conference on Privacy,
  Security and Trust (PST), IEEE, 2018, pp. 1--3.

\bibitem{mousavi2019use}
S.~Mousavi, F.~Afghah, J.~D. Ashdown, K.~Turck, Use of a quantum genetic
  algorithm for coalition formation in large-scale uav networks, Ad Hoc
  Networks 87 (2019) 26--36.

\bibitem{capkun2003self}
S.~Capkun, L.~Butty{\'a}n, J.-P. Hubaux, {Self-organized Public-key Management
  for Mobile Ad hoc Networks}, IEEE Transactions on mobile computing 2~(1)
  (2003) 52--64.

\bibitem{zhang2016fakemask}
L.~Zhang, Z.~Cai, X.~Wang, {Fakemask: A Novel Privacy Preserving Approach for
  Smartphones}, IEEE Transactions on Network and Service Management 13~(2)
  (2016) 335--348.

\bibitem{andrea2015internet}
I.~Andrea, C.~Chrysostomou, G.~Hadjichristofi, {Internet of Things: Security
  Vulnerabilities and Challenges}, in: {Computers and Communication (ISCC),
  2015 IEEE Symposium on}, IEEE, 2015, pp. 180--187.

\bibitem{eschenauer2002key}
L.~Eschenauer, V.~D. Gligor, {A key-management Scheme for Distributed Sensor
  Networks}, in: Proceedings of the 9th ACM conference on Computer and
  communications security, ACM, 2002, pp. 41--47.

\bibitem{keshavarz2020real}
M.~Keshavarz, A.~Shamsoshoara, F.~Afghah, J.~Ashdown, A real-time framework for
  trust monitoring in a network of unmanned aerial vehicles, in: IEEE INFOCOM
  2020-IEEE Conference on Computer Communications Workshops (INFOCOM WKSHPS),
  IEEE, 2020, pp. 677--682.

\bibitem{pongle2015survey}
P.~Pongle, G.~Chavan, {A survey: Attacks on RPL and 6LoWPAN in IoT}, in:
  {Pervasive Computing (ICPC), 2015 International Conference on}, IEEE, 2015,
  pp. 1--6.

\bibitem{han2015practical}
J.~Han, M.~Ha, D.~Kim, {Practical Security Analysis for the Constrained Node
  Networks: Focusing on the DTLS Protocol}, in: 2015 5th International
  Conference on the Internet of Things (IoT), IEEE, 2015, pp. 22--29.

\bibitem{delavar2017puf}
M.~Delavar, S.~Mirzakuchaki, M.~H. Ameri, J.~Mohajeri, {PUF-based solutions for
  secure communications in Advanced Metering Infrastructure (AMI)},
  International Journal of Communication Systems 30~(9) (2017) e3195.

\bibitem{valdivieso2014sdn}
A.~L. Valdivieso~Caraguay, A.~Benito~Peral, L.~I. Barona~Lopez, L.~J.
  Garcia~Villalba, {SDN: Evolution and Opportunities in the Development IoT
  Applications}, International Journal of Distributed Sensor Networks 10~(5)
  (2014) 735142.

\bibitem{moosavi2015sea}
S.~R. Moosavi, T.~N. Gia, A.-M. Rahmani, E.~Nigussie, S.~Virtanen, J.~Isoaho,
  H.~Tenhunen, {SEA: a Secure and Efficient Authentication and Authorization
  Architecture for IoT-based Healthcare using Smart Gateways}, Procedia
  Computer Science 52 (2015) 452--459.

\bibitem{gaur2015smart}
A.~Gaur, B.~Scotney, G.~Parr, S.~McClean, {Smart City Architecture and its
  Applications based on IoT}, Procedia computer science 52 (2015) 1089--1094.

\bibitem{vuvcinic2015oscar}
M.~Vu{\v{c}}ini{\'c}, B.~Tourancheau, F.~Rousseau, A.~Duda, L.~Damon,
  R.~Guizzetti, {OSCAR: Object Security Architecture for the Internet of
  Things}, Ad Hoc Networks 32 (2015) 3--16.

\bibitem{chakrabarty2016secure}
S.~Chakrabarty, D.~W. Engels, {A Secure IoT Architecture for Smart Cities}, in:
  {2016 13th IEEE annual consumer communications \& networking conference
  (CCNC)}, IEEE, 2016, pp. 812--813.

\bibitem{gubbi2013internet}
J.~Gubbi, R.~Buyya, S.~Marusic, M.~Palaniswami, {Internet of Things (IoT): A
  vision, Architectural Elements, and Future Directions}, Future generation
  computer systems 29~(7) (2013) 1645--1660.

\bibitem{rahimi2018security}
H.~Rahimi, A.~Zibaeenejad, P.~Rajabzadeh, A.~A. Safavi, {On the Security of the
  5G-IoT Architecture}, in: Proceedings of the international conference on
  smart cities and internet of things, ACM, 2018, p.~10.

\bibitem{chen2011novel}
D.~Chen, G.~Chang, L.~Jin, X.~Ren, J.~Li, F.~Li, {A Novel Secure Architecture
  for the Internet of Things}, in: {Genetic and Evolutionary Computing (ICGEC),
  2011 Fifth International Conference on}, IEEE, 2011, pp. 311--314.

\bibitem{nawir2016internet}
M.~Nawir, A.~Amir, N.~Yaakob, O.~B. Lynn, {Internet of Things (IoT): Taxonomy
  of Security Attacks}, in: {Electronic Design (ICED), 2016 3rd International
  Conference on}, IEEE, 2016, pp. 321--326.

\bibitem{li2015internet}
S.~Li, L.~Da~Xu, S.~Zhao, {The Internet of Things: a Survey}, Information
  Systems Frontiers 17~(2) (2015) 243--259.

\bibitem{trappe2015low}
W.~Trappe, R.~Howard, R.~S. Moore, {Low-energy Security: Limits and
  Opportunities in the Internet of Things}, IEEE Security \& Privacy 13~(1)
  (2015) 14--21.

\bibitem{alsaadi2015internet}
E.~Alsaadi, A.~Tubaishat, {Internet of Things: Features, Challenges, and
  Vulnerabilities}, International Journal of Advanced Computer Science and
  Information Technology 4~(1) (2015) 1--13.

\bibitem{belapurkar2009distributed}
A.~Belapurkar, A.~Chakrabarti, H.~Ponnapalli, N.~Varadarajan, S.~Padmanabhuni,
  S.~Sundarrajan, {Distributed Systems Security: Issues, Processes and
  Solutions}, John Wiley \& Sons, 2009.

\bibitem{sushma2011security}
D.~N. Sushma, V.~Nandal, {Security Threats in Wireless Sensor Networks}, IJCSMS
  International Journal of Computer Science \& Management Studies 11~(01)
  (2011) 59--63.

\bibitem{zhang2014sybil}
K.~Zhang, X.~Liang, R.~Lu, X.~Shen, {Sybil Attacks and their Defenses in the
  Internet of Things}, IEEE Internet of Things Journal 1~(5) (2014) 372--383.

\bibitem{yilmaz2015survey}
M.~H. Y{\i}lmaz, H.~Arslan, {A survey: Spoofing Attacks in Physical Layer
  Security}, in: Local Computer Networks Conference Workshops (LCN Workshops),
  2015 IEEE 40th, IEEE, 2015, pp. 812--817.

\bibitem{hossain2015towards}
M.~M. Hossain, M.~Fotouhi, R.~Hasan, {Towards an Analysis of Security Issues,
  Challenges, and Open Problems in the Internet of Things}, in: {Services
  (SERVICES), 2015 IEEE World Congress on}, IEEE, 2015, pp. 21--28.

\bibitem{alam2014analysis}
S.~Alam, D.~De, {Analysis of Security Threats in Wireless Sensor Networks},
  arXiv preprint arXiv:1406.0298.

\bibitem{mayzaud2016taxonomy}
A.~Mayzaud, R.~Badonnel, I.~Chrisment, {A Taxonomy of Attacks in RPL-based
  Internet of Things}, International Journal of Network Security 18~(3) (2016)
  459--473.

\bibitem{sabeel2013categorized}
U.~Sabeel, S.~Maqbool, {Categorized Security Threats in the Wireless Sensor
  Networks: Countermeasures and Security Management Schemes}, International
  Journal of Computer Applications 64~(16).

\bibitem{mosenia2017comprehensive}
A.~Mosenia, N.~K. Jha, {A Comprehensive Study of Security of
  Internet-of-Things}, IEEE Transactions on Emerging Topics in Computing 5~(4)
  (2017) 586--602.

\bibitem{li2007optimal}
M.~Li, I.~Koutsopoulos, R.~Poovendran, {Optimal Jamming Attacks and Network
  Defense Policies in Wireless Sensor Networks}, in: INFOCOM 2007. 26th IEEE
  International Conference on Computer Communications. IEEE, IEEE, 2007, pp.
  1307--1315.

\bibitem{mpitziopoulos2009survey}
A.~Mpitziopoulos, D.~Gavalas, C.~Konstantopoulos, G.~Pantziou, {A survey on
  Jamming Attacks and Countermeasures in WSNs}, IEEE Communications Surveys \&
  Tutorials 11~(4).

\bibitem{becher2006tampering}
A.~Becher, Z.~Benenson, M.~Dornseif, {Tampering with Motes: Real-world Physical
  Attacks on Wireless Sensor Networks}, in: {International Conference on
  Security in Pervasive Computing}, Springer, 2006, pp. 104--118.

\bibitem{lemke2006embedded}
K.~Lemke, {Embedded Security: Physical Protection against Tampering Attacks},
  in: {Embedded Security in Cars}, Springer, 2006, pp. 207--217.

\bibitem{schramm2003new}
K.~Schramm, T.~Wollinger, C.~Paar, {A New Class of Collision Attacks and its
  Application to DES}, in: {International Workshop on Fast Software
  Encryption}, Springer, 2003, pp. 206--222.

\bibitem{bogdanov2008multiple}
A.~Bogdanov, {Multiple-differential Side-channel Collision Attacks on AES}, in:
  {International Workshop on Cryptographic Hardware and Embedded Systems},
  Springer, 2008, pp. 30--44.

\bibitem{sanadhya2008new}
S.~K. Sanadhya, P.~Sarkar, {New Collision Attacks against up to 24-step SHA-2},
  in: {International conference on cryptology in India}, Springer, 2008, pp.
  91--103.

\bibitem{tarouco2012internet}
L.~M.~R. Tarouco, L.~M. Bertholdo, L.~Z. Granville, L.~M.~R. Arbiza,
  F.~Carbone, M.~Marotta, J.~J.~C. de~Santanna, {Internet of Things in
  Healthcare: Interoperatibility and Security Issues}, in: {Communications
  (ICC), 2012 IEEE International Conference on}, IEEE, 2012, pp. 6121--6125.

\bibitem{heer2011security}
T.~Heer, O.~Garcia-Morchon, R.~Hummen, S.~L. Keoh, S.~S. Kumar, K.~Wehrle,
  {Security Challenges in the IP-based Internet of Things}, Wireless Personal
  Communications 61~(3) (2011) 527--542.

\bibitem{varga2017security}
P.~Varga, S.~Plosz, G.~Soos, C.~Hegedus, {Security Threats and Issues in
  Automation IoT}, in: {Factory Communication Systems (WFCS), 2017 IEEE 13th
  International Workshop on}, IEEE, 2017, pp. 1--6.

\bibitem{burhanuddin2018review}
M.~Burhanuddin, A.~A.-J. Mohammed, R.~Ismail, M.~E. Hameed, A.~N. Kareem,
  H.~Basiron, {A Review on Security Challenges and Features in Wireless Sensor
  Networks: IoT Perspective}, Journal of Telecommunication, Electronic and
  Computer Engineering (JTEC) 10~(1-7) (2018) 17--21.

\bibitem{liu2012authentication}
J.~Liu, Y.~Xiao, C.~P. Chen, {Authentication and Access Control in the Internet
  of Things}, in: {Distributed Computing Systems Workshops (ICDCSW), 2012 32nd
  International Conference on}, IEEE, 2012, pp. 588--592.

\bibitem{brachmann2012end}
M.~Brachmann, S.~L. Keoh, O.~G. Morchon, S.~S. Kumar, {End-to-end Transport
  Security in the IP-based Internet of Things}, in: {Computer Communications
  and Networks (ICCCN), 2012 21st International Conference on}, IEEE, 2012, pp.
  1--5.

\bibitem{bysani2011survey}
L.~K. Bysani, A.~K. Turuk, {A survey on Selective Forwarding Attack in Wireless
  Sensor Networks}, in: {Devices and Communications (ICDeCom), 2011
  International Conference on}, IEEE, 2011, pp. 1--5.

\bibitem{khan2011comprehensive}
W.~Z. Khan, X.~Yang, M.~Y. Aalsalem, Q.~Arshad, {Comprehensive Study of
  Selective Forwarding Attack in Wireless Sensor Networks}, International
  Journal of Computer Network and Information Security 3~(1) (2011) 1.

\bibitem{karlof2003secure}
C.~Karlof, D.~Wagner, {Secure Routing in Sensor Networks: Attacks and
  Countermeasures} (2003).

\bibitem{krontiris2008launching}
I.~Krontiris, T.~Giannetsos, T.~Dimitriou, {Launching a Sinkhole Attack in
  Wireless Sensor Networks; the Intruder Side}, in: {IEEE International
  Conference on Wireless \& Mobile Computing, Networking \& Communication},
  IEEE, 2008, pp. 526--531.

\bibitem{choi2009sinkhole}
B.~G. Choi, E.~J. Cho, J.~H. Kim, C.~S. Hong, J.~H. Kim, {A Sinkhole Attack
  Detection Mechanism for LQI based Mesh Routing in WSN}, in: {Information
  Networking, 2009. ICOIN 2009. International Conference on}, IEEE, 2009, pp.
  1--5.

\bibitem{gandhewar2012detection}
N.~Gandhewar, R.~Patel, {Detection and Prevention of Sinkhole Attack on AODV
  Protocol in Mobile Adhoc Network}, in: {Computational Intelligence and
  Communication Networks (CICN), 2012 Fourth International Conference on},
  IEEE, 2012, pp. 714--718.

\bibitem{borgohain2015survey}
T.~Borgohain, U.~Kumar, S.~Sanyal, {Survey of security and privacy issues of
  Internet of Things}, arXiv preprint arXiv:1501.02211.

\bibitem{senie1998network}
D.~Senie, P.~Ferguson, {Network Ingress Filtering: Defeating Denial of Service
  Attacks which employ IP Source Address Spoofing}, Network.

\bibitem{hamid2006routing}
M.~A. Hamid, M.~Rashid, C.~S. Hong, {Routing Security in Sensor Network: Hello
  Flood Attack and Defense}, IEEE ICNEWS (2006) 2--4.

\bibitem{singh2010hello}
V.~P. Singh, S.~Jain, J.~Singhai, {Hello Flood Attack and its Countermeasures
  in Wireless Sensor Networks}, International Journal of Computer Science
  Issues (IJCSI) 7~(3) (2010) 23.

\bibitem{sharma2010wireless}
K.~Sharma, M.~Ghose, {Wireless Sensor networks: An overview on its Security
  Threats}, IJCA, Special Issue on “Mobile Ad-hoc Networks” MANETs (2010)
  42--45.

\bibitem{win2008analysis}
K.~S. Win, {Analysis of Detecting Wormhole Attack in Wireless Networks}, in:
  {World Academy of Science, Engineering and Technology}, Citeseer, 2008, pp.
  422--428.

\bibitem{jhaveri2010manet}
R.~H. Jhaveri, A.~D. Patel, J.~D. Parmar, B.~I. Shah, et~al., {MANET Routing
  Protocols and Wormhole Attack against AODV}, International Journal of
  Computer Science and Network Security 10~(4) (2010) 12--18.

\bibitem{wood2002denial}
A.~D. Wood, J.~A. Stankovic, {Denial of Service in Sensor Networks}, computer
  35~(10) (2002) 54--62.

\bibitem{yi2005resisting}
P.~Yi, Z.~Dai, Y.~Zhong, S.~Zhang, {Resisting Flooding Attacks in Ad hoc
  Networks}, in: {Information technology: Coding and computing, 2005. ITCC
  2005. International conference on}, Vol.~2, IEEE, 2005, pp. 657--662.

\bibitem{eddy2007tcp}
W.~Eddy, {TCP SYN Flooding Attacks and Common Mitigations}, Tech. rep., RFC
  (2007).

\bibitem{joncheray1995simple}
L.~Joncheray, {A Simple Active Attack Against TCP}, in: {USENIX Security
  Symposium}, 1995, pp. 2--15.

\bibitem{roosta2006taxonomy}
T.~Roosta, S.~Shieh, S.~Sastry, {Taxonomy of Security Attacks in Sensor
  Networks and Countermeasures}, in: {The first IEEE international conference
  on system integration and reliability improvements}, Vol.~25, 2006, p.~94.

\bibitem{pathan2006security}
A.-S.~K. Pathan, H.-W. Lee, C.~S. Hong, {Security in Wireless Sensor Networks:
  Issues and Challenges}, in: {Advanced Communication Technology, 2006. ICACT
  2006. The 8th International Conference}, Vol.~2, IEEE, 2006, pp. 6--pp.

\bibitem{murdoch2006hot}
S.~J. Murdoch, {Hot or not: Revealing Hidden Services by their Clock Skew}, in:
  {Proceedings of the 13th ACM conference on Computer and communications
  security}, ACM, 2006, pp. 27--36.

\bibitem{manzo2005time}
M.~Manzo, T.~Roosta, S.~Sastry, {Time Synchronization Attacks in Sensor
  Networks}, in: {Proceedings of the 3rd ACM workshop on Security of ad hoc and
  sensor networks}, ACM, 2005, pp. 107--116.

\bibitem{arackaparambil2010reliability}
C.~Arackaparambil, S.~Bratus, A.~Shubina, D.~Kotz, {On the Reliability of
  Wireless Fingerprinting using Clock Skews}, in: {Proceedings of the third ACM
  conference on Wireless network security}, ACM, 2010, pp. 169--174.

\bibitem{yu2006detecting}
B.~Yu, B.~Xiao, {Detecting Selective Forwarding Attacks in Wireless Sensor
  Networks}, in: {Parallel and distributed processing symposium, 2006. IPDPS
  2006. 20th international}, IEEE, 2006, pp. 8--pp.

\bibitem{roy2006attack}
S.~Roy, S.~Setia, S.~Jajodia, {Attack-resilient Hierarchical Data Aggregation
  in Sensor Networks}, in: {Proceedings of the fourth ACM workshop on Security
  of ad hoc and sensor networks}, ACM, 2006, pp. 71--82.

\bibitem{rezvani2015secure}
M.~Rezvani, A.~Ignjatovic, E.~Bertino, S.~Jha, {Secure Data Aggregation
  Technique for Wireless Sensor Networks in the presence of Collusion Attacks},
  IEEE transactions on Dependable and Secure Computing 12~(1) (2015) 98--110.

\bibitem{atamli2014threat}
A.~W. Atamli, A.~Martin, {Threat-based Security Analysis for the Internet of
  Things}, in: 2014 International Workshop on Secure Internet of Things, IEEE,
  2014, pp. 35--43.

\bibitem{lu2018internet}
Y.~Lu, L.~Da~Xu, {Internet of Things (IoT) Cybersecurity Research: a Review of
  Current Research Topics}, IEEE Internet of Things Journal 6~(2) (2018)
  2103--2115.

\bibitem{lin2016iot}
H.~Lin, N.~Bergmann, {IoT Privacy and Security Challenges for Smart Home
  Environments}, Information 7~(3) (2016) 44.

\bibitem{baccelli2018riot}
E.~Baccelli, C.~G{\"u}ndo{\u{g}}an, O.~Hahm, P.~Kietzmann, M.~S. Lenders,
  H.~Petersen, K.~Schleiser, T.~C. Schmidt, M.~W{\"a}hlisch, {RIOT: An Open
  Source Operating System for Low-end Embedded Devices in the IoT}, IEEE
  Internet of Things Journal 5~(6) (2018) 4428--4440.

\bibitem{keshavarz2019automatic}
M.~Keshavarz, M.~Anwar, The automatic detection of sensitive data in smart
  homes, in: International Conference on Human-Computer Interaction, Springer,
  2019, pp. 404--416.

\bibitem{gassend2002controlled}
B.~Gassend, D.~Clarke, M.~Van~Dijk, S.~Devadas, {Controlled physical random
  functions}, in: 18th Annual Computer Security Applications Conference, 2002.
  Proceedings., IEEE, 2002, pp. 149--160.

\bibitem{tehranipoor2011introduction}
M.~Tehranipoor, C.~Wang, {Introduction to hardware security and trust},
  Springer Science \& Business Media, 2011.

\bibitem{eckmann2002statl}
S.~T. Eckmann, G.~Vigna, R.~A. Kemmerer, {STATL: An Attack Language for
  State-based Intrusion Detection}, Journal of computer security 10~(1-2)
  (2002) 71--103.

\bibitem{ning2012technology}
H.~Ning, S.~Hu, {Technology Classification, Industry, and Education for Future
  Internet of Things}, International Journal of Communication Systems 25~(9)
  (2012) 1230--1241.

\bibitem{yang2012multi}
X.~Yang, Z.~Li, Z.~Geng, H.~Zhang, {A Multi-layer Security Model for Internet
  of Things}, in: Internet of things, Springer, 2012, pp. 388--393.

\bibitem{song2013security}
Y.~Song, {Security in Internet of Things} (2013).

\bibitem{leach2005universally}
P.~Leach, M.~Mealling, R.~Salz, {A Universally Unique Identifier (UUID) URN
  Namespace}, Tech. rep., Network Working Group (2005).

\bibitem{song2010semantic}
Z.~Song, A.~A. C{\'a}rdenas, R.~Masuoka, {Semantic Middleware for the Internet
  of Things}, in: Internet of Things (IOT), 2010, IEEE, 2010, pp. 1--8.

\bibitem{shelby2014constrained}
Z.~Shelby, K.~Hartke, C.~Bormann, {The Constrained Application Protocol
  (CoAP)}, Tech. rep., IETF (2014).

\bibitem{hunkeler2008mqtt}
U.~Hunkeler, H.~L. Truong, A.~Stanford-Clark, {MQTT-S—A Publish/subscribe
  Protocol for Wireless Sensor Networks}, in: {Communication systems software
  and middleware and workshops, 2008. comsware 2008. 3rd international
  conference on}, IEEE, 2008, pp. 791--798.

\bibitem{khan2012future}
R.~Khan, S.~U. Khan, R.~Zaheer, S.~Khan, {Future Internet: the Internet of
  Things Architecture, Possible Applications and Key Challenges}, in: Frontiers
  of Information Technology (FIT), 2012 10th International Conference on, IEEE,
  2012, pp. 257--260.

\bibitem{mousavi2020han}
S.~Mousavi, F.~Afghah, U.~R. Acharya, Han-ecg: An interpretable atrial
  fibrillation detection model using hierarchical attention networks, arXiv
  preprint arXiv:2002.05262.

\bibitem{Koeberl}
P.~Koeberl, {\"U}.~Kocaba{\c{s}}, A.-R. Sadeghi, Memristor pufs: a new
  generation of memory-based physically unclonable functions, in: 2013 Design,
  Automation \& Test in Europe Conference \& Exhibition (DATE), IEEE, 2013, pp.
  428--431.

\bibitem{Holcomb}
D.~E. Holcomb, W.~P. Burleson, K.~Fu,
  \href{https://doi.org/10.1109/TC.2008.212}{Power-up sram state as an
  identifying fingerprint and source of true random numbers}, IEEE Trans.
  Comput. 58~(9) (2009) 1198–1210.
\newblock \href {http://dx.doi.org/10.1109/TC.2008.212}
  {\path{doi:10.1109/TC.2008.212}}.
\newline\urlprefix\url{https://doi.org/10.1109/TC.2008.212}

\bibitem{Sutar}
S.~{Sutar}, A.~{Raha}, V.~{Raghunathan}, Memory-based combination pufs for
  device authentication in embedded systems, IEEE Transactions on Multi-Scale
  Computing Systems 4~(4) (2018) 793--810.

\bibitem{Keller}
C.~{Keller}, F.~{Gürkaynak}, H.~{Kaeslin}, N.~{Felber}, Dynamic memory-based
  physically unclonable function for the generation of unique identifiers and
  true random numbers, in: 2014 IEEE International Symposium on Circuits and
  Systems (ISCAS), 2014, pp. 2740--2743.

\bibitem{sonar2014survey}
K.~Sonar, H.~Upadhyay, {A survey: DDOS Attack on Internet of Things},
  International Journal of Engineering Research and Development 10~(11) (2014)
  58--63.

\bibitem{kumar2016security}
S.~A. Kumar, T.~Vealey, H.~Srivastava, {Security in Internet of Things:
  Challenges, Solutions and Future Directions}, in: {2016 49th Hawaii
  International Conference on System Sciences (HICSS)}, IEEE, 2016, pp.
  5772--5781.

\bibitem{nastase2017security}
L.~Nastase, {Security in the Internet of Things: A survey on Application Layer
  Protocols}, in: 2017 21st International Conference on Control Systems and
  Computer Science (CSCS), IEEE, 2017, pp. 659--666.

\bibitem{rostami2014primer}
M.~Rostami, F.~Koushanfar, R.~Karri, {A Primer on Hardware Security: Models,
  Methods, and Metrics}, Proceedings of the IEEE 102~(8) (2014) 1283--1295.

\bibitem{rostami2013hardware}
M.~Rostami, F.~Koushanfar, J.~Rajendran, R.~Karri, {Hardware Security: Threat
  Models and Metrics}, in: Proceedings of the International Conference on
  Computer-Aided Design, IEEE Press, 2013, pp. 819--823.

\bibitem{koushanfar2012can}
F.~Koushanfar, S.~Fazzari, C.~McCants, W.~Bryson, P.~Song, M.~Sale,
  M.~Potkonjak, {Can EDA Combat the Rise of Electronic Counterfeiting?}, in:
  DAC Design Automation Conference 2012, IEEE, 2012, pp. 133--138.

\bibitem{rohatgi2009improved}
P.~Rohatgi, {Improved Techniques for Side-channel Analysis}, in: Cryptographic
  Engineering, Springer, 2009, pp. 381--406.

\bibitem{christofpaar2010understanding}
J.~P. ChristofPaar, B.~Preneel, {Understanding Cryptography: A Textbook for
  Students and ractitioners}, Springer.

\bibitem{mahajan2013study}
P.~Mahajan, A.~Sachdeva, {A study of Encryption Algorithms AES, DES and RSA for
  Security}, Global Journal of Computer Science and Technology.

\bibitem{RSAcrypt55}
Wikipedia, {RSA (cryptosystem) - Wikipedia},
  \url{https://en.wikipedia.org/wiki/RSA\_(cryptosystem)}, (Accessed on
  04/26/2019) (April 2019).

\bibitem{rohatgi2009electromagnetic}
P.~Rohatgi, {Electromagnetic Attacks and Countermeasures}, in: Cryptographic
  Engineering, Springer, 2009, pp. 407--430.

\bibitem{schlosser2013simple}
A.~Schl{\"o}sser, D.~Nedospasov, J.~Kr{\"a}mer, S.~Orlic, J.-P. Seifert,
  {Simple Photonic Emission Analysis of AES}, Journal of Cryptographic
  Engineering 3~(1) (2013) 3--15.

\bibitem{genkin2014rsa}
D.~Genkin, A.~Shamir, E.~Tromer, {RSA Key Extraction via Low-bandwidth Acoustic
  Cryptanalysis}, in: Annual Cryptology Conference, Springer, 2014, pp.
  444--461.

\bibitem{beckmann2009hardware}
N.~Beckmann, M.~Potkonjak, {Hardware-based public-key cryptography with public
  physically unclonable functions}, in: {International Workshop on Information
  Hiding}, Springer, 2009, pp. 206--220.

\bibitem{bi2016enhanced}
Y.~Bi, {Enhanced Hardware Security Using Charge-Based Emerging Device
  Technology}, University of Central Florida, Thesis in Ph.D.

\bibitem{torrance2011state}
R.~Torrance, D.~James, {The State-of-the-art in Semiconductor Reverse
  Rngineering}, in: 2011 48th ACM/EDAC/IEEE Design Automation Conference (DAC),
  IEEE, 2011, pp. 333--338.

\bibitem{saeed2017towards}
S.~M. Saeed, X.~Cui, R.~Wille, A.~Zulehner, K.~Wu, R.~Drechsler, R.~Karri,
  {Towards Reverse Engineering Reversible Logic}, arXiv preprint
  arXiv:1704.08397.

\bibitem{wendt2014hardware}
J.~B. Wendt, M.~Potkonjak, {Hardware obfuscation using PUF-based logic}, in:
  {2014 IEEE/ACM International Conference on Computer-Aided Design (ICCAD)},
  IEEE, 2014, pp. 270--271.

\bibitem{roy2010ending}
J.~A. Roy, F.~Koushanfar, I.~L. Markov, {Ending Piracy of Integrated Circuits},
  Computer 43~(10) (2010) 30--38.

\bibitem{anderson2010puf}
J.~H. Anderson, {A PUF design for secure FPGA-based embedded systems}, in: 2010
  15th Asia and South Pacific Design Automation Conference (ASP-DAC), IEEE,
  2010, pp. 1--6.

\bibitem{karri2010trustworthy}
R.~Karri, J.~Rajendran, K.~Rosenfeld, M.~Tehranipoor, {Trustworthy Hardware:
  Identifying and Classifying Hardware Trojans}, Computer 43~(10) (2010)
  39--46.

\bibitem{tehranipoor2010survey}
M.~Tehranipoor, F.~Koushanfar, {A Survey of Hardware Trojan Taxonomy and
  Detection}, IEEE design \& test of computers 27~(1) (2010) 10--25.

\bibitem{waksman2011silencing}
A.~Waksman, S.~Sethumadhavan, {Silencing Hardware Backdoors}, in: 2011 IEEE
  Symposium on Security and Privacy, IEEE, 2011, pp. 49--63.

\bibitem{dupuis2014novel}
S.~Dupuis, P.-S. Ba, G.~Di~Natale, M.-L. Flottes, B.~Rouzeyre, {A novel
  hardware logic encryption technique for thwarting illegal overproduction and
  hardware Trojans}, in: {2014 IEEE 20th International On-Line Testing
  Symposium (IOLTS)}, IEEE, 2014, pp. 49--54.

\bibitem{venugopalan2018surveying}
V.~Venugopalan, C.~D. Patterson, {Surveying the Hardware Trojan Threat
  Landscape for the Internet-of-Things}, Journal of Hardware and Systems
  Security 2~(2) (2018) 131--141.

\bibitem{jimgreen17:online}
CISCO, {jim\_green\_cisco\_connect.pdf},
  \url{https://www.cisco.com/c/dam/\\global/en\_ph/assets/ciscoconnect/pdf/bigdata/jim\_green\\\_cisco\_connect.pdf},
  (Accessed on 03/26/2020) (2014).

\bibitem{Microsof90:online}
CISCO, {Microsoft Word - IoT Reference Model White Paper June 4, 2014.doc},
  \url{http://cdn.iotwf.com/resources/71/IoT\_Reference\_Model\_White\\\_Paper\_June\_4\_2014.pdf},
  (Accessed on 03/26/2020) (2014).

\bibitem{lesjak2015hardware}
C.~Lesjak, D.~Hein, J.~Winter, {Hardware-security Technologies for Industrial
  IoT: TrustZone and Security Controller}, in: {IECON 2015-41st Annual
  Conference of the IEEE Industrial Electronics Society}, IEEE, 2015, pp.
  002589--002595.

\bibitem{trustzone_link}
{Arm TrustZone Technology},
  \url{https://developer.arm.com/ip-products/\\security-ip/trustzone},
  accessed: 2019-12-04.

\bibitem{cortex-A_link}
Technologies trustzone for cortex-a,
  \url{https://www.arm.com/why-arm/\\technologies/trustzone-for-cortex-a},
  accessed: 2019-12-04.

\bibitem{cortex-M_link}
Technologies trustzone for cortex-m,
  \url{https://www.arm.com/why-arm/\\technologies/trustzone-for-cortex-m},
  accessed: 2019-12-04.

\bibitem{trustzone_link_arm}
{TrustZone Technology for Armv8-M},
  \url{https://developer.arm.com\\/ip-products/security-ip/trustzone/trustzone-for-cortex-m},
  accessed: 2020-01-13.

\bibitem{ruhrmair2012security}
U.~R{\"u}hrmair, S.~Devadas, F.~Koushanfar, {Security based on Physical
  Unclonability and Disorder}, in: {Introduction to Hardware Security and
  Trust}, Springer, 2012, pp. 65--102.

\bibitem{puftutorial}
C.~Herder, M.-D. Yu, F.~Koushanfar, S.~Devadas, {Physical Unclonable Functions
  and Applications: A Tutorial}, Proceedings of the IEEE 102~(8) (2014)
  1126--1141.

\bibitem{maiti2012robust}
A.~Maiti, I.~Kim, P.~Schaumont, {A Robust Physical Unclonable Function with
  Enhanced Challenge-response Set}, IEEE Transactions on Information Forensics
  and Security 7~(1) (2012) 333--345.

\bibitem{bautista2016survey}
I.~A. Bautista~Adames, J.~Das, S.~Bhanja, {Survey of Emerging Technology based
  Physical Unclonable Funtions}, in: {Proceedings of the 26th edition on Great
  Lakes Symposium on VLSI}, ACM, 2016, pp. 317--322.

\bibitem{korenda2018secret}
A.~R. Korenda, F.~Afghah, B.~Cambou, {A Secret Key Generation Scheme for
  Internet of Things using Ternary-states ReRAM-based Physical Unclonable
  Functions}, in: 2018 14th International Wireless Communications \& Mobile
  Computing Conference (IWCMC), IEEE, 2018, pp. 1261--1266.

\bibitem{cambouafghah2015}
B.~Cambou, F.~Afghah, {Physically Unclonable Functions with Multi-states and
  Machine Learning}, in: {14th International Workshop on Cryptographic
  Architectures Embedded in Logic Devices (CryptArchi)}, 2016, pp. 1--1.

\bibitem{ruhrmair2010strong}
U.~R{\"u}hrmair, H.~Busch, S.~Katzenbeisser, Strong pufs: models,
  constructions, and security proofs, in: Towards hardware-intrinsic security,
  Springer, 2010, pp. 79--96.

\bibitem{pappu2002physical}
R.~Pappu, B.~Recht, J.~Taylor, N.~Gershenfeld, {Physical One-way Functions},
  Science 297~(5589) (2002) 2026--2030.

\bibitem{tuyls2006rfid}
P.~Tuyls, L.~Batina, {RFID-tags for Anti-counterfeiting}, in: Cryptographers’
  Track at the RSA Conference, Springer, 2006, pp. 115--131.

\bibitem{shamsoshoara2019ring}
A.~Shamsoshoara, {Ring Oscillator and its Application as Physical Unclonable
  Function (PUF) for Password Management}, arXiv preprint arXiv:1901.06733.

\bibitem{vijayakumar2017improving}
A.~Vijayakumar, V.~Patil, S.~Kundu, {On Improving Reliability of SRAM-based
  Physically Unclonable Functions}, Journal of Low Power Electronics and
  Applications 7~(1) (2017) 2.

\bibitem{bohm2011microcontroller}
C.~B{\"o}hm, M.~Hofer, W.~Pribyl, {A Microcontroller SRAM-PUF}, in: {Network
  and System Security (NSS), 2011 5th International Conference on}, IEEE, 2011,
  pp. 269--273.

\bibitem{holcomb2007initial}
D.~E. Holcomb, W.~P. Burleson, K.~Fu, et~al., {Initial SRAM State as a
  Fingerprint and Source of True Random Numbers for RFID Tags}, in:
  {Proceedings of the Conference on RFID Security}, Vol.~7, 2007, p.~2.

\bibitem{ReRAMternaryPUF}
B.~Cambou, M.~Orlowski, {PUF designed with Resistive RAM and Ternary States},
  in: {Proceedings of the 11th Annual Cyber and Information Security Research
  Conference}, ACM, 2016, p.~1.

\bibitem{helfmeier2014physical}
C.~Helfmeier, C.~Boit, D.~Nedospasov, S.~Tajik, J.-P. Seifert, {Physical
  Vulnerabilities of Physically Unclonable Functions}, in: {Proceedings of the
  conference on Design, Automation \& Test in Europe}, European Design and
  Automation Association, 2014, p. 350.

\bibitem{cambou2019encoding}
B.~Cambou, Encoding data for cells in a puf that corresponds to a response in a
  challenge response pair, uS Patent 10,439,828 (Oct.~8 2019).

\bibitem{pavan1997flash}
P.~Pavan, R.~Bez, P.~Olivo, E.~Zanoni, {Flash Memory Cells-an Overview},
  Proceedings of the IEEE 85~(8) (1997) 1248--1271.

\bibitem{chang2007endurance}
Y.-H. Chang, J.-W. Hsieh, T.-W. Kuo, {Endurance Enhancement of Flash-memory
  Storage Systems: an Afficient Static Wear Leveling Design}, in: {Proceedings
  of the 44th annual Design Automation Conference}, ACM, 2007, pp. 212--217.

\bibitem{akinaga2010resistive}
H.~Akinaga, H.~Shima, {Resistive Random Access Memory (ReRAM) based on Metal
  Oxides}, Proceedings of the IEEE 98~(12) (2010) 2237--2251.

\bibitem{tehrani2006status}
S.~Tehrani, {Status and Outlook of MRAM Memory Technology}, in: {Electron
  Devices Meeting, 2006. IEDM'06. International}, IEEE, 2006, pp. 1--4.

\bibitem{puf-cprng2020}
S.~{Kalanadhabhatta}, D.~{Kumar}, K.~K. {Anumandla}, S.~A. {Reddy},
  A.~{Acharyya}, {PUF-Based Secure Chaotic Random Number Generator Design
  Methodology}, IEEE Transactions on Very Large Scale Integration (VLSI)
  Systems (2020) 1--5.

\bibitem{wang2019puf}
Q.~Wang, M.~Gao, G.~Qu, {PUF-PassSE: A PUF based Password Strength Enhancer for
  IoT Applications}, in: {20th International Symposium on Quality Electronic
  Design (ISQED)}, IEEE, 2019, pp. 198--203.

\bibitem{cambou2018addressable}
B.~Cambou, {Addressable PUF generators for database-free password management
  system} (2018).

\bibitem{cambou2020secure}
B.~F. Cambou, {Secure digital signatures using physical unclonable function
  devices with reduced error rates}, uS Patent App. 16/560,502 (Mar.~5 2020).

\bibitem{cambou2019data}
B.~Cambou, {Data compiler for true random number generation and related
  methods}, uS Patent 10,175,949 (Jan.~8 2019).

\bibitem{tuyls2006read}
P.~Tuyls, G.-J. Schrijen, B.~{\v{S}}kori{\'c}, J.~Van~Geloven, N.~Verhaegh,
  R.~Wolters, {Read-proof Hardware from Protective Coatings}, in: International
  Workshop on Cryptographic Hardware and Embedded Systems, Springer, 2006, pp.
  369--383.

\bibitem{skoric2007experimental}
B.~Skoric, G.-J. Schrijen, W.~Ophey, R.~Wolters, N.~Verhaegh, J.~van Geloven,
  {Experimental Hardware for Coating PUFs and Optical PUFs}, in: Security with
  Noisy Data, Springer, 2007, pp. 255--268.

\bibitem{skoric2006information}
B.~Skoric, S.~Maubach, T.~A. Kevenaar, P.~Tuyls, {Information-theoretic
  Analysis of Coating PUFs}, IACR Cryptology ePrint Archive 2006 (2006) 101.

\bibitem{kursawe2009reconfigurable}
K.~Kursawe, A.-R. Sadeghi, D.~Schellekens, B.~Skoric, P.~Tuyls, {Reconfigurable
  Physical Unclonable Functions-enabling Technology for Tamper-resistant
  Storage}, in: {Hardware-Oriented Security and Trust, 2009. HOST'09. IEEE
  International Workshop on}, IEEE, 2009, pp. 22--29.

\bibitem{ruhrmair2013optical}
U.~R{\"u}hrmair, C.~Hilgers, S.~Urban, A.~Weiersh{\"a}user, E.~Dinter,
  B.~Forster, C.~Jirauschek, {Optical PUFs Reloaded}, Eprint. Iacr. Org.

\bibitem{suh2007physical}
G.~E. Suh, S.~Devadas, {Physical Unclonable Functions for Device Authentication
  and Secret Key Generation}, in: {Proceedings of the 44th annual design
  automation conference}, ACM, 2007, pp. 9--14.

\bibitem{morozov2010analysis}
S.~Morozov, A.~Maiti, P.~Schaumont, {An Analysis of Delay based PUF
  Implementations on FPGA}, in: {International Symposium on Applied
  Reconfigurable Computing}, Springer, 2010, pp. 382--387.

\bibitem{fruhashi2011arbiter}
K.~Fruhashi, M.~Shiozaki, A.~Fukushima, T.~Murayama, T.~Fujino, {The
  Arbiter-PUF with High Uniqueness utilizing Novel Arbiter Circuit with
  Delay-Time Measurement}, in: {Circuits and Systems (ISCAS), 2011 IEEE
  International Symposium on}, IEEE, 2011, pp. 2325--2328.

\bibitem{maiti2009improving}
A.~Maiti, P.~Schaumont, {Improving the Quality of a Physical Unclonable
  Function using Configurable Ring Oscillators}, in: Field Programmable Logic
  and Applications, 2009. FPL 2009. International Conference on, IEEE, 2009,
  pp. 703--707.

\bibitem{yin2009temperature}
C.-E. Yin, G.~Qu, {Temperature-aware Cooperative Ring Oscillator PUF}, in:
  {Hardware-Oriented Security and Trust, 2009. HOST'09. IEEE International
  Workshop on}, IEEE, 2009, pp. 36--42.

\bibitem{cambou2016puf}
B.~Cambou, M.~Orlowski, {PUF Designed with Resistive RAM and Ternary States},
  in: {Proceedings of the 11th Annual Cyber and Information Security Research
  Conference}, ACM, 2016, p.~1.

\bibitem{afghah2018reram}
F.~Afghah, B.~Cambou, M.~Abedini, S.~Zeadally, {A ReRAM Physically Unclonable
  Function (ReRAM PUF)-based Approach to Enhance Authentication Security in
  Software Defined Wireless Networks}, International Journal of Wireless
  Information Networks (2018) 1--13.

\bibitem{kumar2008butterfly}
S.~S. Kumar, J.~Guajardo, R.~Maes, G.-J. Schrijen, P.~Tuyls, {The Butterfly PUF
  protecting IP on every FPGA}, in: {Hardware-Oriented Security and Trust,
  2008. HOST 2008. IEEE International Workshop on}, IEEE, 2008, pp. 67--70.

\bibitem{ruhrmair2014pufs}
U.~R{\"u}hrmair, D.~E. Holcomb, {PUFs at a glance}, in: {Proceedings of the
  conference on Design, Automation \& Test in Europe}, European Design and
  Automation Association, 2014, p. 347.

\bibitem{guajardo2007fpga}
J.~Guajardo, S.~S. Kumar, G.-J. Schrijen, P.~Tuyls, {FPGA Intrinsic PUFs and
  their use for IP Protection}, in: International workshop on cryptographic
  hardware and embedded systems, Springer, 2007, pp. 63--80.

\bibitem{guajardo2007physical}
J.~Guajardo, S.~S. Kumar, G.-J. Schrijen, P.~Tuyls, {Physical Unclonable
  Functions and Public-key Crypto for FPGA IP Protection}, in: Field
  Programmable Logic and Applications, 2007. FPL 2007. International Conference
  on, IEEE, 2007, pp. 189--195.

\bibitem{claes2011comparison}
M.~Claes, V.~van~der Leest, A.~Braeken, {Comparison of SRAM and FF PUF in 65nm
  Technology}, in: Nordic Conference on Secure IT Systems, Springer, 2011, pp.
  47--64.

\bibitem{assiri2019key}
S.~Assiri, B.~Cambou, D.~D. Booher, D.~G. Miandoab, M.~Mohammadinodoushan, {Key
  Exchange using Ternary system to Enhance Security}, in: {2019 IEEE 9th Annual
  Computing and Communication Workshop and Conference (CCWC)}, IEEE, 2019, pp.
  0488--0492.

\bibitem{Howmanyc21_online}
scmo, {How many computers are there in the world? — SCMO},
  \url{https://www.scmo.net/faq/2019/8/9/how-many-compaters-is-there\\
  -in-the-world\#:~:text=In\%202019\%2C\%20there\%20were\%20over,\\
  servers\%2C\%20desktops\%2C\%20and\%20laptops.}, (Accessed on 06/30/2020)
  (August 2019).

\bibitem{koushanfar2012hardware}
F.~Koushanfar, {Hardware metering: A survey}, in: Introduction to Hardware
  Security and Trust, Springer, 2012, pp. 103--122.

\bibitem{zamanzadeh2016higher}
S.~Zamanzadeh, A.~Jahanian, {Higher security of ASIC fabrication process
  against reverse engineering attack using automatic netlist encryption
  methodology}, Microprocessors and Microsystems 42 (2016) 1--9.

\bibitem{rajendran2012security}
J.~Rajendran, Y.~Pino, O.~Sinanoglu, R.~Karri, {Security analysis of logic
  obfuscation}, in: Proceedings of the 49th Annual Design Automation
  Conference, 2012, pp. 83--89.

\bibitem{alkabani2007active}
Y.~Alkabani, F.~Koushanfar, {Active Hardware Metering for Intellectual Property
  Protection and Security.}, in: USENIX security symposium, 2007, pp. 291--306.

\bibitem{khaleghi2018hardware}
S.~Khaleghi, W.~Rao, {Hardware obfuscation using strong PUFs}, in: 2018 IEEE
  Computer Society Annual Symposium on VLSI (ISVLSI), IEEE, 2018, pp. 321--326.

\bibitem{suh2005}
G.~E. Suh, C.~W. O'Donnell, S.~Devadas, {AEGIS: A Single-chip Secure
  Processor}, Information Security Technical Report 10~(2) (2005) 63--73.

\bibitem{kang2014performance}
H.~Kang, Y.~Hori, T.~Katashita, M.~Hagiwara, K.~Iwamura, {Performance Analysis
  for PUF Data using Fuzzy Extractor}, in: {Ubiquitous Information Technologies
  and Applications}, Springer, 2014, pp. 277--284.

\bibitem{ziola2014authentication}
T.~Ziola, Z.~Paral, S.~Devadas, G.~E. Suh, V.~Khandelwal, {Authentication with
  Physical Unclonable Functions}, uS Patent 8,782,396 (Jul.~15 2014).

\bibitem{vskoric2005robust}
B.~{\v{S}}kori{\'c}, P.~Tuyls, W.~Ophey, {Robust Key Extraction from Physical
  Uncloneable Functions}, in: {International Conference on Applied Cryptography
  and Network Security}, Springer, 2005, pp. 407--422.

\bibitem{lim2005extracting}
D.~Lim, J.~W. Lee, B.~Gassend, G.~E. Suh, M.~Van~Dijk, S.~Devadas, {Extracting
  Secret Keys from Integrated Circuits}, IEEE Transactions on Very Large Scale
  Integration (VLSI) Systems 13~(10) (2005) 1200--1205.

\bibitem{maes2012pufky}
R.~Maes, A.~Van~Herrewege, I.~Verbauwhede, {PUFKY: A Fully Functional PUF-based
  Cryptographic Key Generator}, Cryptographic Hardware and Embedded
  Systems--CHES 2012 (2012) 302--319.

\bibitem{bosch2008efficient}
C.~B{\"o}sch, J.~Guajardo, A.-R. Sadeghi, J.~Shokrollahi, P.~Tuyls, {Efficient
  Helper Data Key Extractor on FPGAs}, in: {International Workshop on
  Cryptographic Hardware and Embedded Systems}, Springer, 2008, pp. 181--197.

\bibitem{maes2009low}
R.~Maes, P.~Tuyls, I.~Verbauwhede, {Low-Overhead Implementation of a Soft
  Decision Helper Data Algorithm for SRAM PUFs.}, in: {CHES}, Vol.~9, Springer,
  2009, pp. 332--347.

\bibitem{puchinger2015error}
S.~Puchinger, S.~M{\"u}elich, M.~Bossert, M.~Hiller, G.~Sigl, {On Error
  Correction for Physical Unclonable Functions}, in: {SCC 2015; 10th
  International ITG Conference on Systems, Communications and Coding;
  Proceedings of}, VDE, 2015, pp. 1--6.

\bibitem{chen2017high}
B.~Chen, T.~Ignatenko, F.~M. Willems, R.~Maes, E.~van~der Sluis, G.~Selimis,
  {High-Rate Error Correction Schemes for SRAM-PUFs based on Polar Codes},
  arXiv preprint arXiv:1701.07320.

\bibitem{Korenda_SECON}
A.~R. Korenda, F.~Afghah, B.~Cambou, C.~Philabaum, {A Proof of Concept
  SRAM-based Physically Unclonable Function (PUF) Key Generation Mechanism for
  IoT Devices}, in: 2019 16th Annual IEEE International Conference on Sensing,
  Communication, and Networking (SECON), IEEE, 2019, pp. 1--8.

\bibitem{merli2011side}
D.~Merli, D.~Schuster, F.~Stumpf, G.~Sigl, {Side-channel analysis of PUFs and
  fuzzy extractors}, in: International Conference on Trust and Trustworthy
  Computing, Springer, 2011, pp. 33--47.

\bibitem{delvaux2014key}
J.~Delvaux, I.~Verbauwhede, {Key-recovery attacks on various RO PUF
  constructions via helper data manipulation}, in: Proceedings of the
  conference on Design, Automation \& Test in Europe, European Design and
  Automation Association, 2014, p.~72.

\bibitem{yin2010lisa}
C.-E.~D. Yin, G.~Qu, {LISA: Maximizing RO PUF's secret extraction}, in: 2010
  IEEE International Symposium on Hardware-Oriented Security and Trust (HOST),
  IEEE, 2010, pp. 100--105.

\bibitem{yin2013design}
C.-E. Yin, G.~Qu, Q.~Zhou, {Design and implementation of a group-based RO PUF},
  in: 2013 Design, Automation \& Test in Europe Conference \& Exhibition
  (DATE), IEEE, 2013, pp. 416--421.

\bibitem{yin2013improving}
C.-E. Yin, G.~Qu, {Improving PUF security with regression-based distiller}, in:
  Proceedings of the 50th Annual Design Automation Conference, 2013, pp. 1--6.

\bibitem{delvaux2014helper}
J.~Delvaux, D.~Gu, D.~Schellekens, I.~Verbauwhede, {Helper data algorithms for
  PUF-based key generation: Overview and analysis}, IEEE Transactions on
  Computer-Aided Design of Integrated Circuits and Systems 34~(6) (2014)
  889--902.

\bibitem{delvaux2014attacking}
J.~Delvaux, I.~Verbauwhede, {Attacking PUF-based pattern matching key
  generators via helper data manipulation}, in: {Cryptographers’ Track at the
  RSA Conference}, Springer, 2014, pp. 106--131.

\bibitem{paral2011reliable}
Z.~Paral, S.~Devadas, {Reliable and efficient PUF-based key generation using
  pattern matching}, in: {2011 IEEE International Symposium on
  Hardware-Oriented Security and Trust}, IEEE, 2011, pp. 128--133.

\end{thebibliography}

\end{document}